\documentclass[useAMS,usenatbib]{mn2e}
\usepackage{graphicx}

\newcommand{\Ion}[2]{#1{\,\sc#2}}
\newcommand{\SDSS}[1]{SDSS\,J{#1}}
\newcommand{\kms}{\mbox{$\mathrm{km\,s^{-1}}$}}
\newcommand{\MSUN}{\mbox{$\mathrm{M_{\odot}}$}}

\newcommand{\msy}{\mbox{$\mathrm{M_{\odot}\,yr^{-1}}$}}

\title[Eclipsing PCEBs from the CRTS]{Eclipsing Post Common Envelope Binaries from the Catalina Surveys}

\author[S. G. Parsons et al.]{S.~G.~Parsons$^{1,2}$\thanks{steven.parsons@uv.cl},
B.~T.~G{\"a}nsicke$^{1}$,
T.~R.~Marsh$^{1}$,
A.~J.~Drake$^{3}$,
V.~S.~Dhillon$^{4}$,
\newauthor
S.~P.~Littlefair$^{4}$,
S.~Pyrzas$^{1,5}$,
A.~Rebassa-Mansergas$^{2}$
and M.~R.~Schreiber$^{2,6}$
\\
$^{1}$Department of Physics, University of Warwick, Coventry, CV4 7AL\\
$^{2}$Departmento de F{\'i}sica y Astronom{\'i}a, Universidad de
Valpara{\'i}so, Avenida Gran Bretana 1111, Valpara{\'i}so, Chile\\
$^{3}$California Institute of Technology, 1200 E. California Blvd, CA 91225,
USA\\
$^{4}$Department of Physics and Astronomy, University of Sheffield, Sheffield
S3 7RH, UK\\
$^{5}$Instituto de Astronom\'ia, Universidad Cat\'olica del Norte, Avenida 
Angamos 0610, Casilla 1280, Antofagasta, Chile\\
$^{6}$Millennium Nucleus ``Protoplanetary Disks in ALMA Early Science'',
Universidad de Valparaiso, Av. Gran Bretana 111, Chile}
\begin{document}
\input{references.cls}
\date{Accepted 2012 October 31. Received 2012 October 24; in original form 2012 September 11}

\pagerange{\pageref{firstpage}--\pageref{lastpage}} \pubyear{2012}

\maketitle

\label{firstpage}

\begin{abstract}

We analyse the Catalina Real-time Transient Survey light curves of 835 spectroscopically confirmed white dwarf plus main-sequence binaries from the Sloan Digital Sky Survey (SDSS) with $g<19$, in search of new eclipsing systems. We identify 29 eclipsing systems, 12 of which were previously unknown. This brings the total number of eclipsing white dwarf plus main-sequence binaries to 49. Our set of new eclipsing systems contains two with periods of 1.9 and 2.3 days making them the longest period eclipsing white dwarf binaries known. We also identify one system which shows very large ellipsoidal modulation (almost 0.3 magnitudes), implying that the system is both very close to Roche-lobe overflow and at high inclination. However, our follow up photometry failed to firmly detect an eclipse meaning that this system either contains a cool white dwarf and hence the eclipse is very shallow and undetectable in our red-sensitive photometry or that it is non-eclipsing. Radial velocity measurements for the main-sequence stars in three of our newly identified eclipsing systems imply that their white dwarf masses are lower than those inferred from modelling their SDSS spectra. 13 non-eclipsing post common envelope binaries were also identified, from either reflection or ellipsoidal modulation effects. The white dwarfs in our newly discovered eclipsing systems span a wide range of parameters, including; low mass ($\sim$$0.3$\MSUN), very hot (80,000\,K) and a DC white dwarf. The spectral types of the main-sequence stars range from M2 to M6. This makes our sample ideal for testing white dwarf and low-mass star mass-radius relationships as well as close binary evolution.  

\end{abstract}


\section{Introduction}

Around 25\% of main-sequence binary systems have stars that are close enough to each other that they will interact at some point in their evolution \citep{willems04}. This interaction is caused by one or both of the stars filling its Roche lobe and causing material to flow from one star to the other. This process can often lead to a common-envelope (CE) phase. The CE phase gives birth to very close binaries and is thought to lead to the creation of some of the Galaxy's most exotic objects, such as cataclysmic variables (CVs), low-mass X-ray binaries, B-type subdwarfs (sdB stars), double degenerates, short gamma ray burst (GRB) progenitors and millisecond pulsars.

One of the most common outcomes of the CE phase are the close detached white dwarf plus main-sequence binaries, known as Post Common Envelope Binaries (PCEBs). These systems offer a unique opportunity to study close binaries without the added complications of accretion, hence these systems can provide us with superb tests of both the common envelope phase itself and the longer-term angular momentum loss mechanisms that drive the evolution of many interacting binary stars \citep{schreiber03}.

There are now over 2000 known white dwarf plus main-sequence binaries \citep{silvestri06,heller11,morgan12,liu12}, with the largest and most homogeneous catalogue presented by \citet{rebassa07,rebassa10,rebassa12a} using data from the Sloan Digital Sky Survey (SDSS; \citealt{york00,adelman08,abazajian09}). Among these $\sim$$1/3$ are thought to be close PCEBs \citep{schreiber10,rebassa11}. This rise in the discovery rate of PCEBs is reflected in a corresponding rise in the number of eclipsing systems: 30 of the 37 currently known eclipsing PCEBs were identified in the last 3 years. Many of these were identified by observing large radial velocity variations in the SDSS sub-spectra (each SDSS spectrum is the average of typically three 15 minute exposures or {\it sub-spectra}) \citep{nebot09,pyrzas09,pyrzas12} or from searches for pulsations from the white dwarf \citep{steinfadt08}. However, an increasing number of eclipsing systems are now being discovered in large scale time-domain surveys such as the Palomar Transit Factory (PTF) \citep{law11a,law11b}, the multi-epoch SDSS photometric survey (Stripe 82) \citep{becker11}\footnote{Note that only 6 of the 42 candidate white dwarf plus M dwarf binaries in \citet{becker11} show evidence of a white dwarf in their SDSS spectrum. Of these only one (SDSS\,J013851.54-001621.6) has been confirmed as eclipsing \citep{parsons12c}.} and a survey at the Isaac Newton Telescope \citep{almenara12}.

By far the most successful search for eclipsing PCEBs was made by \citet{drake09,drake10} using data from the CSS (Catalina Sky Survey) and the Catalina Real Time Transient Survey (CRTS). They discovered 26 eclipsing systems; 13 were previously unknown eclipsing PCEBs, 6 were previously known eclipsing PCEBs, 3 were eclipsing cataclysmic variables, 3 were sdB+dM eclipsing binaries and 1 turned out to be a double white dwarf eclipsing binary \citep{parsons11b}. The primary aim of that study was to detect transiting planets around white dwarfs, since even Earth-sized planets would produce deep eclipses due to the small size of white dwarfs. Therefore, \citet{drake10} selected their targets from the white dwarf catalogue of \citet{eisenstein06} supplemented with additional photometric objects from the SDSS that were selected using the ($u-g$,$g-r$) colour plane with a cut that included the majority of the \citet{eisenstein06} white dwarfs. This selection rejects PCEBs where the companion star noticeably contributes in the $g$ or $r$ bands, and hence implies that the optical colours of Drake et al's targets are dominated by their white dwarf components. Therefore, the \citet{drake10} sample is heavily biased towards hot white dwarfs with late-type companions.

Here we present a search for eclipsing PCEBs combining the large catalogue of WDMS binaries spectroscopically identified in SDSS \citep{rebassa12a} with the detailed CSS light curves of these objects. We recover all 17 previously known eclipsing PCEBs contained in our target list, and identify 12 additional ones, plus one candidate eclipsing PCEB. The 12 newly discovered eclipsing PCEBs have been followed-up with high-speed photometry.

\section{Data reduction}

We selected all targets from \citet{rebassa12a} with $g<19$, a total of 966 systems. Not all of these systems have been observed as part of the CSS and several targets were highly blended with nearby stars and hence the resultant light curves were very poor. We discarded these systems, resulting in 835 light curves in total. 

The Catalina Sky Survey has been running since mid 2005 and is designed to discover Near-Earth Objects. It uses the 0.7m f/1.9 Catalina Schmidt Telescope with an eight square degree field of view. Full details of the CSS can be found in \citet{drake09}. The observing strategy is to observe each field in a sequence of four 30-second exposures, spaced evenly over approximately 30 minutes, typically reaching V magnitudes of 19 to 20. The CSS dataset consists of fields covered from a few times to more than 400 times. We used data obtained up to November 2011.

In order to improve the calibration of the CSS photometry and hence the associated uncertainties, we decided to perform differential photometry on the reduced (bias-subtracted and flat-fielded) CSS images. This also allowed us to identify images in which the target was not detected (e.g. deeply eclipsing systems).

For each target we produced a series of $10'\times10'$ image cutouts from the CSS data, centred on the target. All of these images contained a substantial number of additional nearby stars to the main target. Since, by definition, all the fields had been observed as part of the SDSS, the additional sources had SDSS magnitudes and could be used to calculate the zeropoint, and hence flux calibrate each frame. We chose to use the SDSS $r$ band magnitudes since this filter most closely approximated the filterless response of the CSS detectors. 

The zero point of each frame was determined by extracting all sources within it using Sextractor \citep{bertin96} and cross-matching them with the SDSS catalogue, we selected all non-blended stars with magnitudes of $15<r<19.5$ and $\delta r<0.05$, as determined by the appropriate flags in casjobs \citep{li08}. We then took the difference between the extracted magnitudes and SDSS magnitudes, removed any values more than $2.5\sigma$ from the mean, and took the median value as the zeropoint for that frame. This corrected for any variations in the observing conditions and also reduced the impact on the zeropoint of any genuinely variable sources in the frame. We also flagged up frames in which the target was not detected. 

\section{Eclipsing systems} \label{sec:eclipsers}

We visually inspected each light curve in order to identify eclipses. This was achieved by identifying any points significantly fainter than the average magnitude of the star ($\sim$2.5$\sigma$ from the mean), or any frames in which the target was not detected. We then inspected the reduced images of each faint point identified to ensure that the target was not on a bad pixel or the edge of a CCD.

\begin{table*}
 \centering
 \begin{minipage}{\textwidth}
 \renewcommand{\thempfootnote}{\fnsymbol{mpfootnote}}
  \centering
  \caption{Identified eclipsing systems. The newly discovered systems are shown in bold and their parameters are taken from \citet{rebassa12a} except for \SDSS{1021+1744}, \SDSS{1028+0931} and \SDSS{1411+1028} where we have radial velocity information and hence are able to constrain the white dwarf masses using the mass function. The white dwarf is not visible spectroscopically in \SDSS{0745+2631}, and \SDSS{1307+2156} contains a featureless DC white dwarf, hence these systems have no parameters listed for their white dwarfs. For the previously discovered systems we list the current best constraints from the literature. References: (1) \citet{pyrzas09}, (2) \citet{parsons10b}, (3) this paper, (4) \citet{drake10}, (5) \citet{parsons12a}, (6) \citet{pyrzas12}, (7) \citet{nebot09}, (8) \citet{parsons12b}.} 
  \label{tab:eclipsers}
  \begin{tabular}{@{}l@{~}llllll@{~}l@{}}
    \hline
    SDSS Name & WD mass & WD $T_\mathrm{eff}$ & MS star  & $r$ mag & Period & T0       & Ref \\
              & (\MSUN) & (K)                & sp type &         & (days) & MJD(BTDB)&           \\
    \hline
    SDSS\,J011009.09+132616.1 & $0.47\pm0.20$ & $25900\pm427$  & M4.0 & 16.86 & 0.332686752(1)     & 53993.949090(2)    & 1,2 \\
    SDSS\,J030308.35+005444.1 & $0.91\pm0.03$ & $<$8000& M4.5 & 18.06 & 0.13443767232(25)  & 53991.1172793(19)  & 1,2 \\
{\bf SDSS\,J074548.63+263123.4}\footnote[2]{Not confirmed as an eclipsing system} & {\bf     } & {\bf      } & {\bf M2.0} & {\bf
17.46} & {\bf 0.2192638284(1)}    & {\bf 53387.2495(10)}     & {\bf 3} \\
{\bf SDSS\,J082145.27+455923.4} & \mbox{\boldmath$0.66\pm0.05$} & \mbox{\boldmath$80938\pm4024$} & {\bf M2.0} & {\bf
  17.52} & {\bf 0.5090912(69)  }    & {\bf 55989.038796(23)}     & {\bf 3} \\
    SDSS\,J083845.86+191416.5 & $0.39\pm0.04$ & $13904\pm424$  & M5.0 & 18.36 & 0.13011225(40)     & 53495.4541(33)     & 4   \\
    SDSS\,J085746.18+034255.3 & $0.514\pm0.049$ & $35300\pm400$  & M8.0 & 18.26 & 0.065096538(3)     & 55552.7127652(8)   & 4,5 \\
    SDSS\,J090812.04+060421.2 & $0.37\pm0.02$ & $17505\pm242$  & M4.0 & 17.28 & 0.1494381329(27)   & 53466.333170(36)   & 4   \\
{\bf SDSS\,J092741.73+332959.1} & \mbox{\boldmath$0.59\pm0.05$} & \mbox{\boldmath$27111\pm494$} & {\bf M3.0} & {\bf
  18.22} & {\bf 2.3082217(65)  }    & {\bf 56074.906137(21)}     & {\bf 3} \\
    SDSS\,J093947.95+325807.3 & $0.52\pm0.03$ & $28389\pm278$  & M4.0 & 18.03 & 0.330989655(21)    & 55587.308823(10)   & 4   \\
{\bf SDSS\,J094634.49+203003.4} & \mbox{\boldmath$0.62\pm0.10$} & \mbox{\boldmath$10307\pm141$} & {\bf M5.0} & {\bf
  18.89} & {\bf 0.2528612195(1)}    & {\bf 56032.945590(25)}     & {\bf 3} \\
    SDSS\,J095719.24+234240.7 & $0.43\pm0.03$ & $25891\pm547$  & M2.0 & 18.06 & 0.150870740(6)     & 55604.830124(6)    & 4   \\
{\bf SDSS\,J095737.59+300136.5} & \mbox{\boldmath$0.42\pm0.05$} & \mbox{\boldmath$28064\pm848$} & {\bf M3.0} & {\bf
  18.78} & {\bf 1.9261278(10)  }    & {\bf 56014.975114(32)}     & {\bf 3} \\
{\bf SDSS\,J102102.25+174439.9} & \mbox{\boldmath$0.50\pm0.05$} & \mbox{\boldmath$32595\pm928$} & {\bf M4.0} & {\bf
  19.01} & {\bf 0.140359073(1)}     & {\bf 56093.90558(12)}     & {\bf 3} \\
{\bf SDSS\,J102857.78+093129.8} & \mbox{\boldmath$0.42\pm0.04$} & \mbox{\boldmath$18756\pm959$} & {\bf M3.0} & {\bf
  15.58} & {\bf 0.235025762(1) }    & {\bf 56001.093511(94)}     & {\bf 3} \\
{\bf SDSS\,J105756.93+130703.5} & \mbox{\boldmath$0.34\pm0.07$} & \mbox{\boldmath$12536\pm978$} & {\bf M5.0} & {\bf
  18.66} & {\bf 0.125162115(23)}    & {\bf 56010.062214(14)}     & {\bf 3} \\
    SDSS\,J121010.13+334722.9 & $0.415\pm0.010$ & $6000\pm200$   & M5.0 & 16.16 & 0.124489764(1)     & 54923.033686(6)    & 6   \\
    SDSS\,J121258.25-012310.2 & $0.439\pm0.002$ & $17707\pm35$  & M4.0 & 16.94 & 0.33587093(13)     & 54104.20917(48)    & 7,8 \\
{\bf SDSS\,J122339.61-005631.1} & \mbox{\boldmath$0.45\pm0.06$} & \mbox{\boldmath$11565\pm59$} & {\bf M6.0} & {\bf
 18.04} & {\bf 0.0900780(13)}      & {\bf 55707.0169865(72)}  & {\bf 3}   \\
    SDSS\,J124432.25+101710.8 & $0.40\pm0.03$ & $21168\pm435$  & M5.0 & 18.34 & 0.2278562(2)       & 53466.3618(11)     & 4   \\
{\bf SDSS\,J130733.49+215636.7} & {\bf     } & {\bf $<$8000   } & {\bf M4.0} & {\bf
  17.42} & {\bf 0.2163221322(1)}    & {\bf 56007.221371(16)}     & {\bf 3} \\
    SDSS\,J132925.21+123025.4 & $0.35\pm0.08$ & $12250\pm1032$  & M8.0 & 17.51 & 0.0809662550(14)   & 55271.05481841(97) & 4   \\
    SDSS\,J134841.61+183410.5 & $0.59\pm0.02$ & $15071\pm167$  & M4.0 & 17.19 & 0.24843148(1)      & 53833.3425(1)      & 4   \\
{\bf SDSS\,J140847.14+295044.9} & \mbox{\boldmath$0.49\pm0.04$} & \mbox{\boldmath$29050\pm484$} & {\bf M5.0} & {\bf
  18.96} & {\bf 0.191790270(24)}    & {\bf 56112.91291(18)}     & {\bf 3} \\
    SDSS\,J141057.73-020236.6 & $0.47\pm0.06$ & $29727\pm508$  & M3.0 & 18.85 & 0.363497(25)       & 53464.4880(36)     & 4   \\
{\bf SDSS\,J141134.70+102839.7} & \mbox{\boldmath$0.36\pm0.04$} & \mbox{\boldmath$30419\pm701$} & {\bf M3.0} & {\bf
  19.13} & {\bf 0.16750990(10) }    & {\bf 56031.172782(48)}     & {\bf 3} \\
    SDSS\,J141536.40+011718.2 & $0.564\pm0.014$ & $55,995\pm673$  & M4.5 & 17.30 & 0.344330838759(92) & 42543.3377143(30)  & 8   \\
    SDSS\,J142355.06+240924.3 & $0.41\pm0.02$ & $32972\pm318$  & M5.0 & 17.87 & 0.38200426(32)     & 53470.39985(18)    & 4   \\
    SDSS\,J143547.87+373338.5 & $0.40\pm0.04$ & $12392\pm328$  & M5.0 & 17.25 & 0.12563114665(67)  & 54148.2035726(35)  & 1   \\
    SDSS\,J145634.30+161137.7 & $0.37\pm0.02$ & $19149\pm262$  & M6.0 & 18.04 & 0.2291202(2)       & 51665.6720(34)     & 4   \\
{\bf SDSS\,J223530.61+142855.0} & \mbox{\boldmath$0.45\pm0.06$} & \mbox{\boldmath$21045\pm711$} & {\bf M4.0} & {\bf
18.83} & {\bf 0.1444564852(34)}   & {\bf 55469.065554(86)}   & {\bf 3}   \\

\hline
\end{tabular}
\vspace{-6mm}
\end{minipage}
\end{table*}

Once we had detected that a system was eclipsing we attempted to determine its period. Initially we calculated a periodogram from the light curve using the \citet{press89} method with inverse variance weights whereby data with smaller errors are given larger weightings. We then folded the light curve on the peak frequency and visually inspected the resultant light curve. In most cases the light curves showed out-of-eclipse variations due to reflection or ellipsoidal modulation effects. In these cases the out-of-eclipse effects allowed us to find the correct period. However, this approach is not ideal for systems that show no out-of-eclipse variations (e.g. longer period systems). The period of these systems could be measured using a box fit similar to those used in exoplanet transit searches \citep[e.g.][]{kovacs02}. However, since we had knowledge of when the system is both in and out of eclipse, we used a simpler (and quicker) approach. We folded the data points over a large range of periods and measured the phase dispersion of the in-eclipse points. We rejected any period in which the in-eclipse points are dispersed by more than 20\% of the orbital period. Furthermore, we insisted that there were no out-of-eclipse points between the in-eclipse points. In all the cases where we used this approach it led to a unique period, this is due to the large number of observations for most targets.

Finally, in order to search for shallower eclipses, we folded all of our light curves on the peak frequency of their periodogram and inspected for eclipses. In cases where we would expect ellipsoidal modulation (e.g. systems with dominant main-sequence star contributions) we also folded the light curves on half the value of the peak frequency, since ellipsoidal modulation causes a double peaked shape in the light curve.

\begin{figure*}
\begin{center}
 \includegraphics[width=0.99\textwidth]{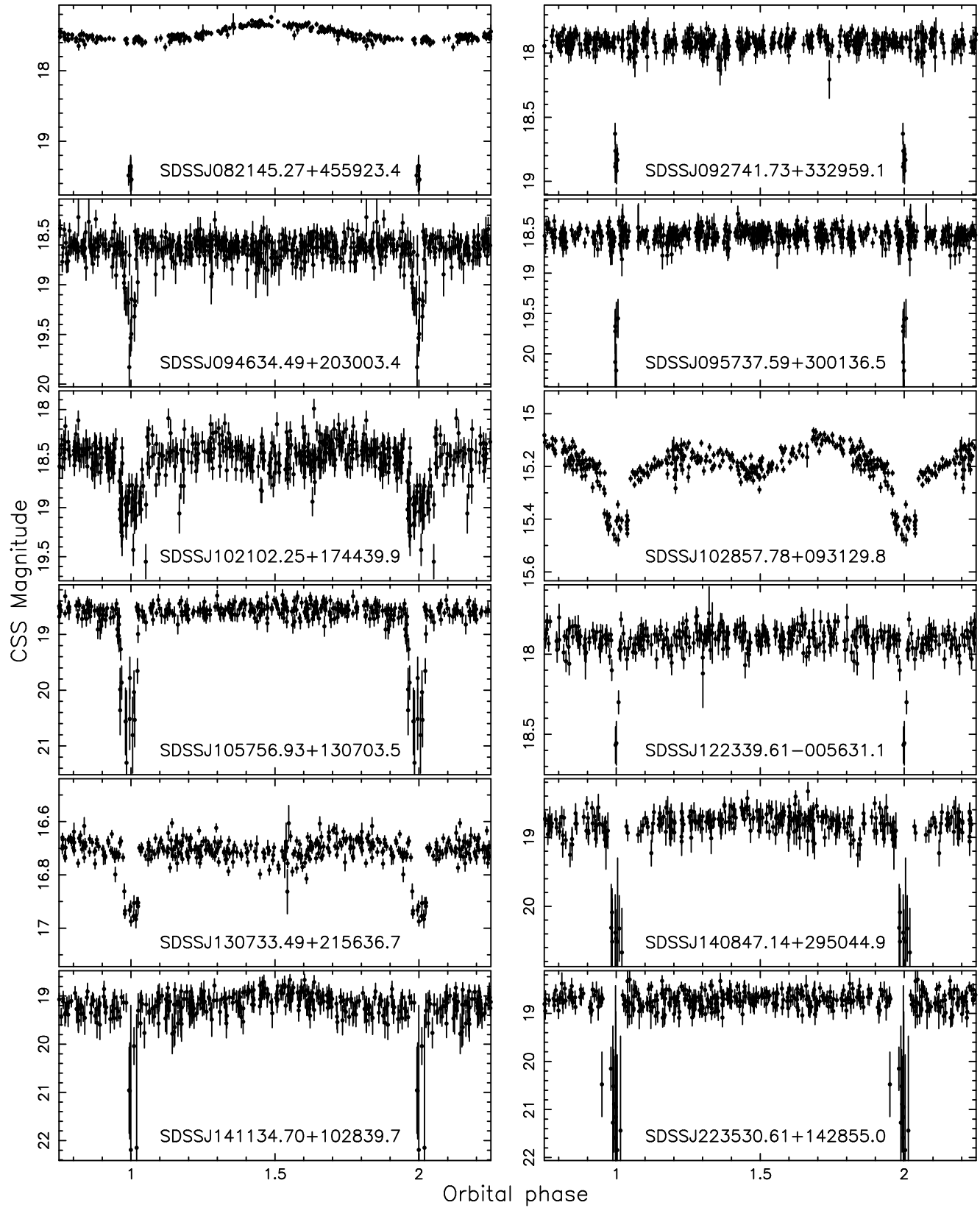}
 \vspace{20mm}
 \caption{Phase-folded CSS light curves of the newly identified eclipsing PCEBs.}
 \label{fig:crts_lcurves}
 \end{center}
\end{figure*}

\begin{figure*}
\begin{center}
 \includegraphics[width=\textwidth]{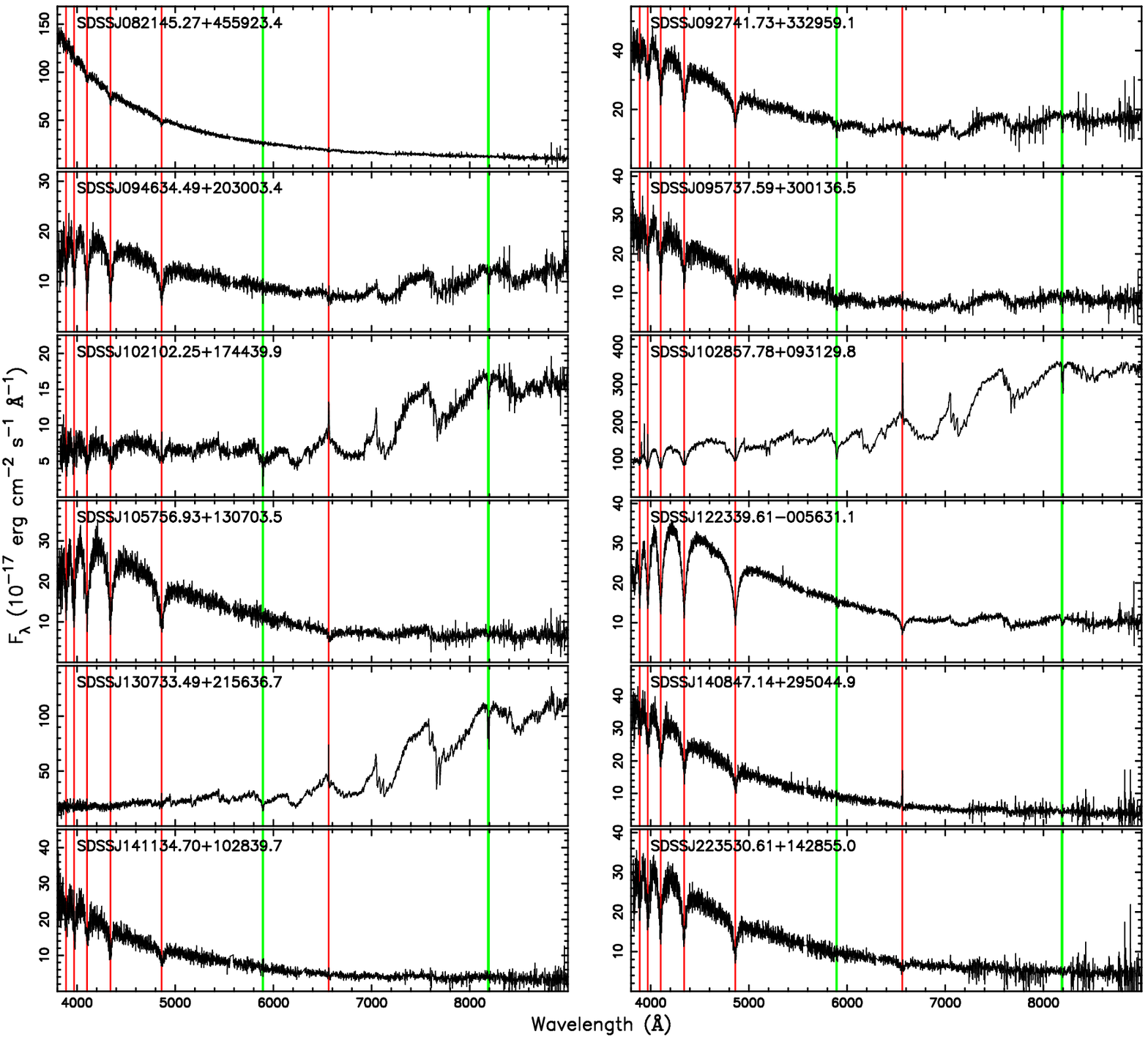}
 \caption{SDSS spectra of the newly identified eclipsing PCEBs. The hydrogen Balmer lines are indicated by red lines (absorption features are from the white dwarf, emission features indicate an active or irradiated main-sequence star) and the \Ion{Na}{i} lines are indicated by the green lines (absorption from the main-sequence star).}
 \label{fig:sdss_specs}
 \end{center}
\end{figure*}

In total we found 29 eclipsing systems, 12 of which were previously unknown, and one candidate eclipsing PCEB which needs better quality photometry for confirmation. This increases the number of confirmed eclipsing PCEBs by more than 20\% from 37 to 49. All our identified eclipsing systems are detailed in Table~\ref{tab:eclipsers}. CSS light curves of the newly identified eclipsing systems are shown in Figure~\ref{fig:crts_lcurves}, and their SDSS spectra are shown in Figure~\ref{fig:sdss_specs}.

\begin{figure}
\includegraphics[angle=-90,width=0.98\columnwidth]{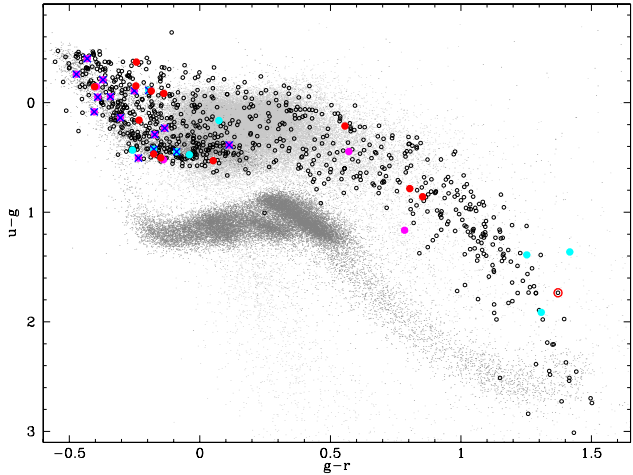}
\caption{Distribution of quasars (light gray dots), stars (dark gray dots), and WDMS binaries (open circles, coloured symbols) in the ($u-g,g-r$) colour plane. All WDMS binaries shown here have SDSS DR7 spectropscopy. Our input target sample (open circles) included 835 WDMS binaries from \citet{rebassa12a} with $g<19$ and good quality CSS light curves. Analysing the CSS light curves of these 835 systems, we identify 29 eclipsing PCEBs of which 17 were previously known (magenta dots) and 12 are new discoveries (red dots). One additional eclipsing PCEB candidate identified here is marked by the red circle. Additional known eclipsing PCEBs that have DR7 SDSS spectra but that were too faint for our magnitude cut are shown as cyan dots. The eclipsing PCEBs announced by \citet{drake09,drake10} are shown by blue crosses.}
  \label{fig:cc_ugr}
\end{figure}

The distribution of all known eclipsing PCEBs with SDSS spectroscopy in the ($u-g,g-r$) colour plane is shown in Figure~\ref{fig:cc_ugr}. This illustrates that our newly discovered eclipsing PCEBs generally contain M-stars with slightly earlier spectral types than the previously known eclipsing systems. This is unsurprising given that the majority of the previously known eclipsing PCEBs in this sample were found by \citet{drake10} in a search for transiting planets around white dwarfs. The colour selection used in the \citet{drake10} study was biased towards systems dominated by the white dwarf, meaning that systems with earlier main-sequence star spectral types were missed. 

\begin{figure}
\includegraphics[width=\columnwidth]{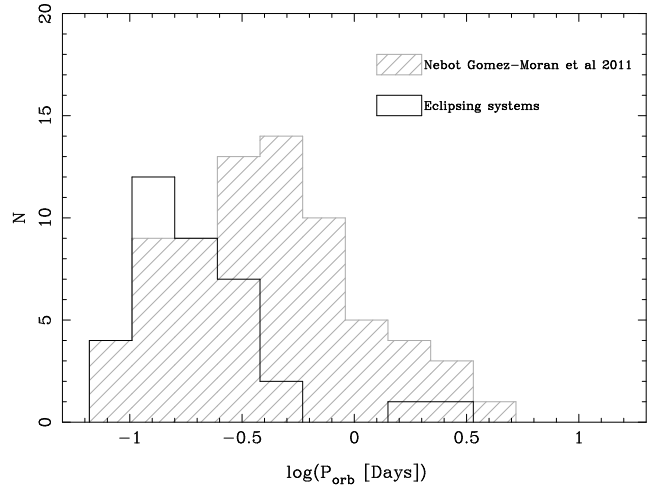}
\caption{Period distribution of all SDSS spectroscopically confirmed eclipsing PCEBs (black) and the orbital period distribution of all SDSS PCEBs from \citet{nebot11} (grey). Since our detection efficiency is high (see Section~\ref{sec:completeness}) the difference between the two distributions mainly reflects the (geometric) probability of a system being eclipsing.}
  \label{fig:pdist}
\end{figure}

The orbital period distribution for all the SDSS eclipsing PCEBs is shown in Figure~\ref{fig:pdist}, as well as the period distribution of all SDSS PCEBs from \citet{nebot11}. Unsurprisingly our systems generally have shorter periods, this is primarily due to the fact that shorter period systems can be eclipsing over a wider range of inclinations. However, we have detected two eclipsing PCEBs with periods in excess of 1.9 days. This is much longer than the previous longest period eclipsing PCEB, V471 Tau, which has a period of only 0.52 days, although, as Figure~\ref{fig:pdist} shows, several non-eclipsing PCEBs have been found with periods this long, or longer \citep{nebot11,rebassa12b}. Our ability to detect these systems is due to the long baseline provided by the Catalina Sky Survey. 

\subsection{System parameters}

The masses, white dwarf temperatures and main-sequence star spectral types were all taken from the catalogue of \citet{rebassa12a}. These were determined by decomposing and fitting the SDSS spectra, see \citet{rebassa07} for a detailed explanation. In brief, the technique first determines the spectral type of the main-sequence star by fitting the SDSS spectrum with a two-component model. The main-sequence component is then subtracted and the residual white dwarf spectrum is fitted with a model grid of white dwarfs from \citet{koester10} to determine its temperature and surface gravity, the mass is then determined using a mass-radius relation for white dwarfs \citep{bergeron95, fontaine01}. This method gives a good first approximation of the stellar parameters, and in most cases gives results consistent with those obtained from high-precision studies (e.g. \citealt{parsons12b}). However, as we will show in Section~\ref{sec:individ}, there are cases in which the deconvolution technique can give erroneous results.

In principle the light curves can also be used to constrain the radii of the two stars and hence the masses via a mass-radius relation. However, the CSS photometry does not sample the white dwarf eclipse sufficently. Our follow-up photometry can only be used to place a lower limit on the size of the main-sequence star ($R_\mathrm{sec}/a$, where $a$ is the orbital separation) and an upper limit on the size of the white dwarf ($R_\mathrm{WD}/a$). Therefore, we adopt the decomposition values for all further discussions.  

\section{follow up photometry} \label{sec:followup}

We obtained follow up high-speed photometry of all our newly identifed eclipsing systems. The majority of these systems were observed with the high-speed camera RISE \citep{steele08} on the Liverpool Telescope (LT). The robotic nature of the LT makes it ideal to observe these systems, particularly the longer period ones. RISE is a frame transfer CCD camera with a single wideband V+R filter and negligible deadtime between frames. We observed one eclipse of each of the newly identified systems using exposure times of between 5 and 25 seconds, depending upon the brightness of the target. The raw data are automatically run through a pipeline that debiases, removes a scaled dark frame and flat-fields the data.  

We also observed two systems, \SDSS{1223-0056} and \SDSS{2235+1428}, with the high-speed camera ULTRACAM \citep{dhillon07}, mounted as a visitor instrument on the New Technology Telescope (NTT) at La Silla.

In all cases the source flux was determined with aperture photometry, using a variable aperture, whereby the radius of the aperture is scaled according to the FWHM, using the ULTRACAM pipeline \citep{dhillon07}. Variations in observing conditions were accounted for by determining the flux relative to nearby comparison stars.

\begin{figure*}
\begin{center}
 \includegraphics[width=\textwidth]{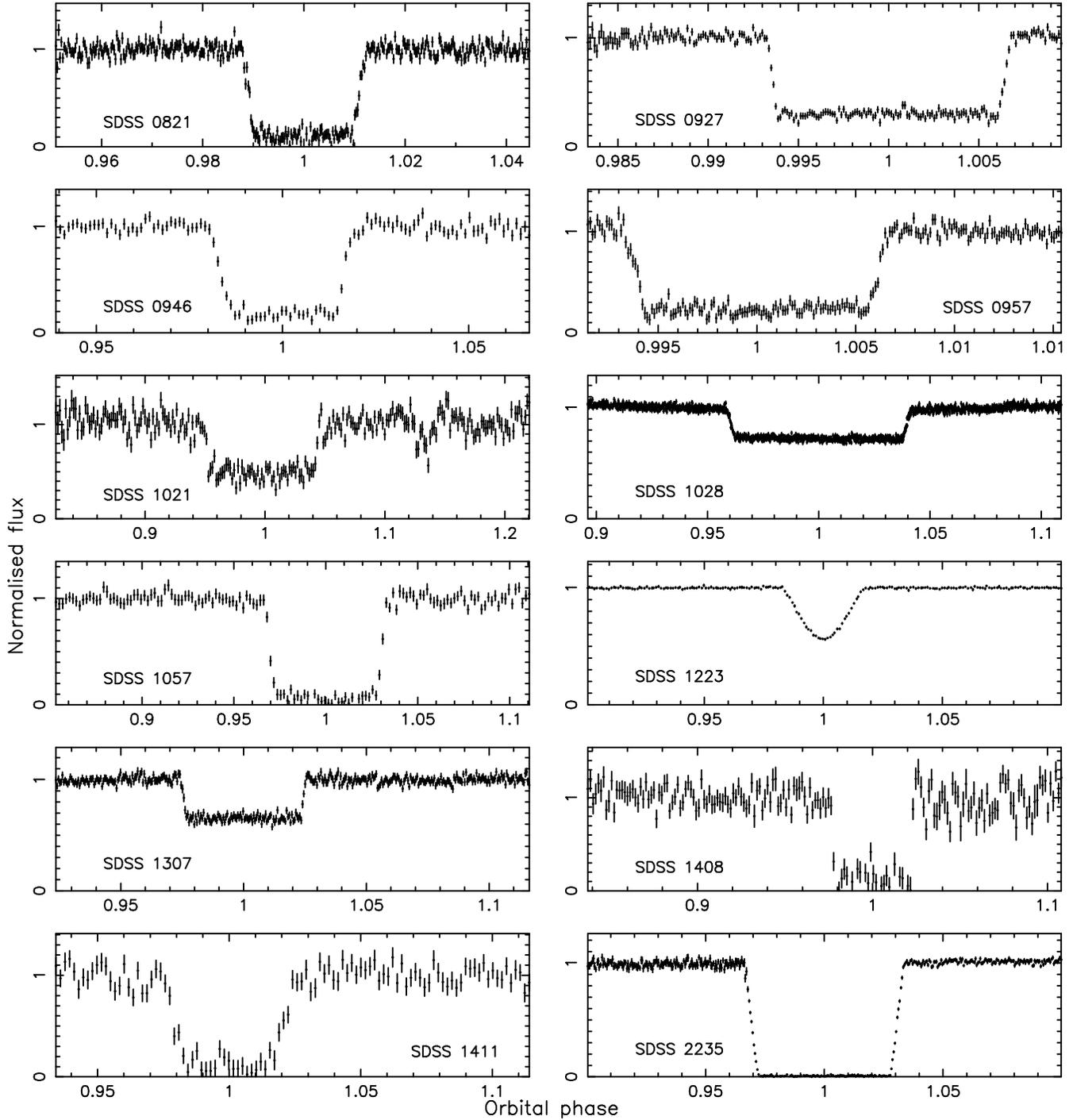}
 \caption{Follow up light curves of the newly identified eclipsing systems. All the data were obtained using RISE on the Liverpool Telescope except for \SDSS{1223-0056} and \SDSS{2235+1428} which were obtained using ULTRACAM on the NTT (the $r$ band eclipses are shown here). The dip seen in the light curve of \SDSS{1021+1744} is likely caused by material ejected from the main-sequence star moving in front of the white dwarf. There also appears to be a flare from the main-sequence star during the egress of the white dwarf.}
 \label{fig:followup}
 \end{center}
\end{figure*}

The follow up light curves of all the confirmed eclipsing systems are shown in Figure~\ref{fig:followup}. We list the mid-eclipse times (T0) from our follow up observations in Table~\ref{tab:eclipsers}.

\section{Notes on individual systems} \label{sec:individ}

\subsubsection*{SDSS J074548.63+263123.4}

\begin{figure*}
\begin{center}
 \includegraphics[width=0.4\textwidth]{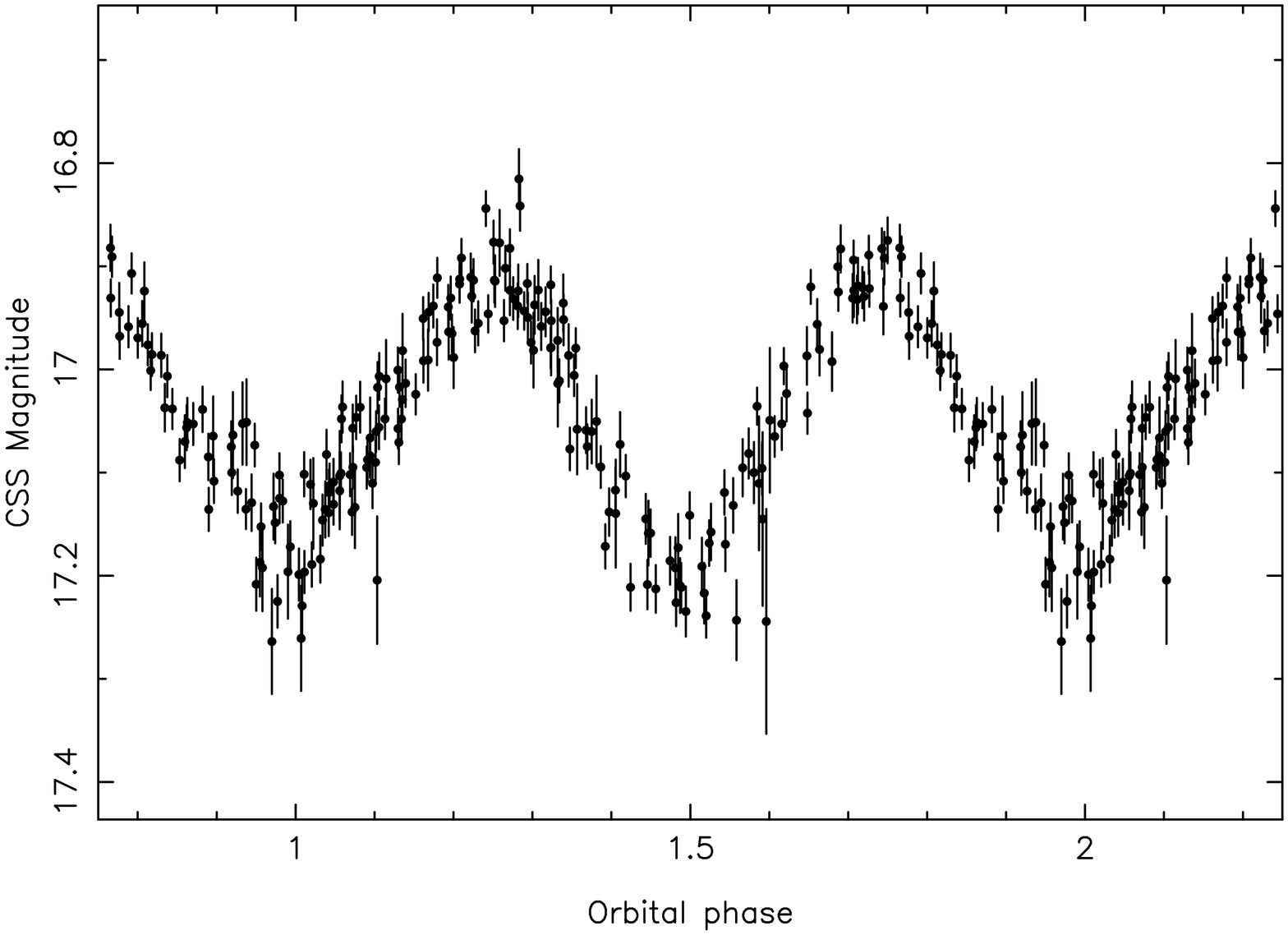}
 \hspace{2mm}
 \includegraphics[width=0.4\textwidth]{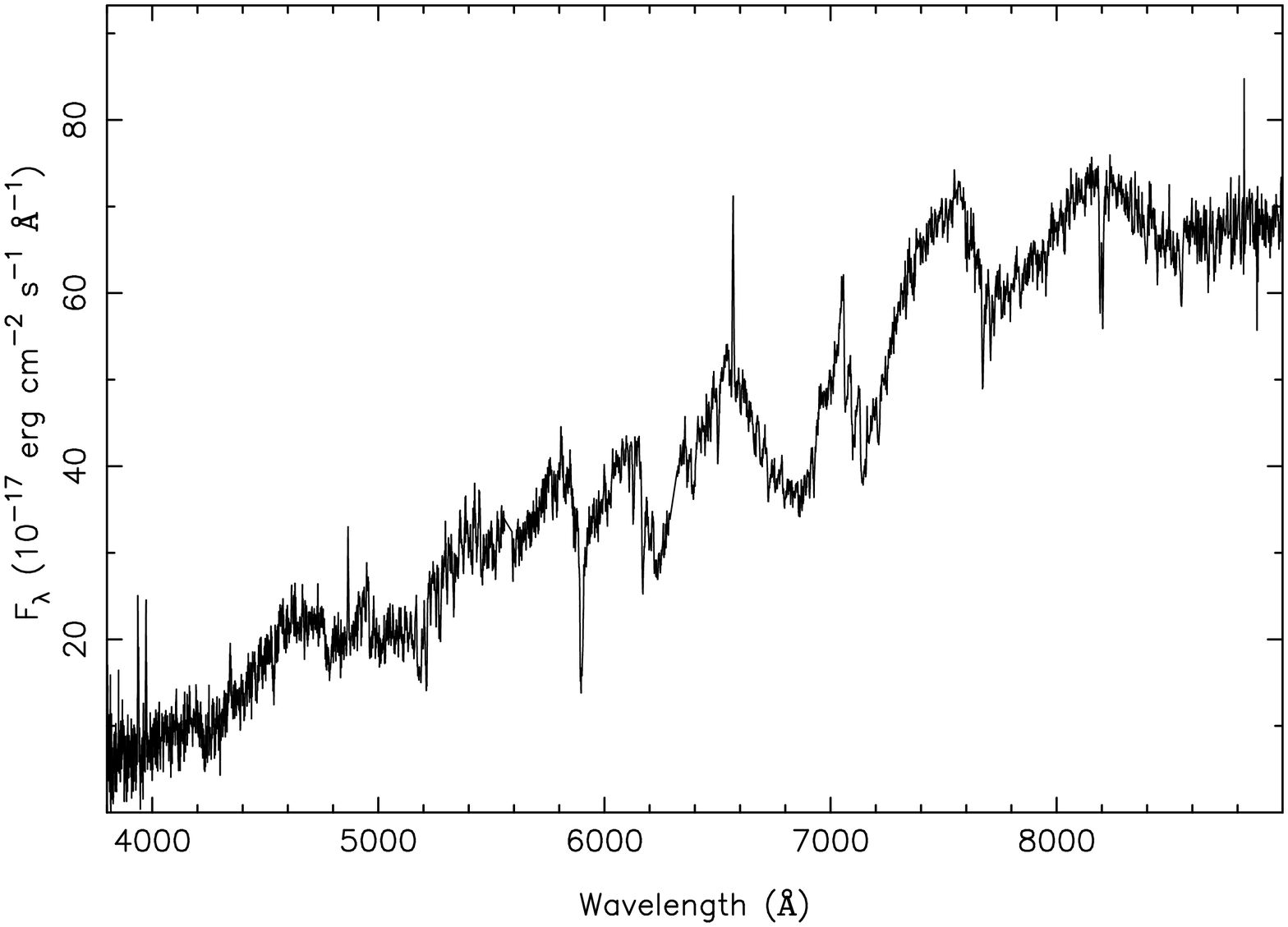}\\
 \vspace{1mm}
 \includegraphics[width=0.4\textwidth]{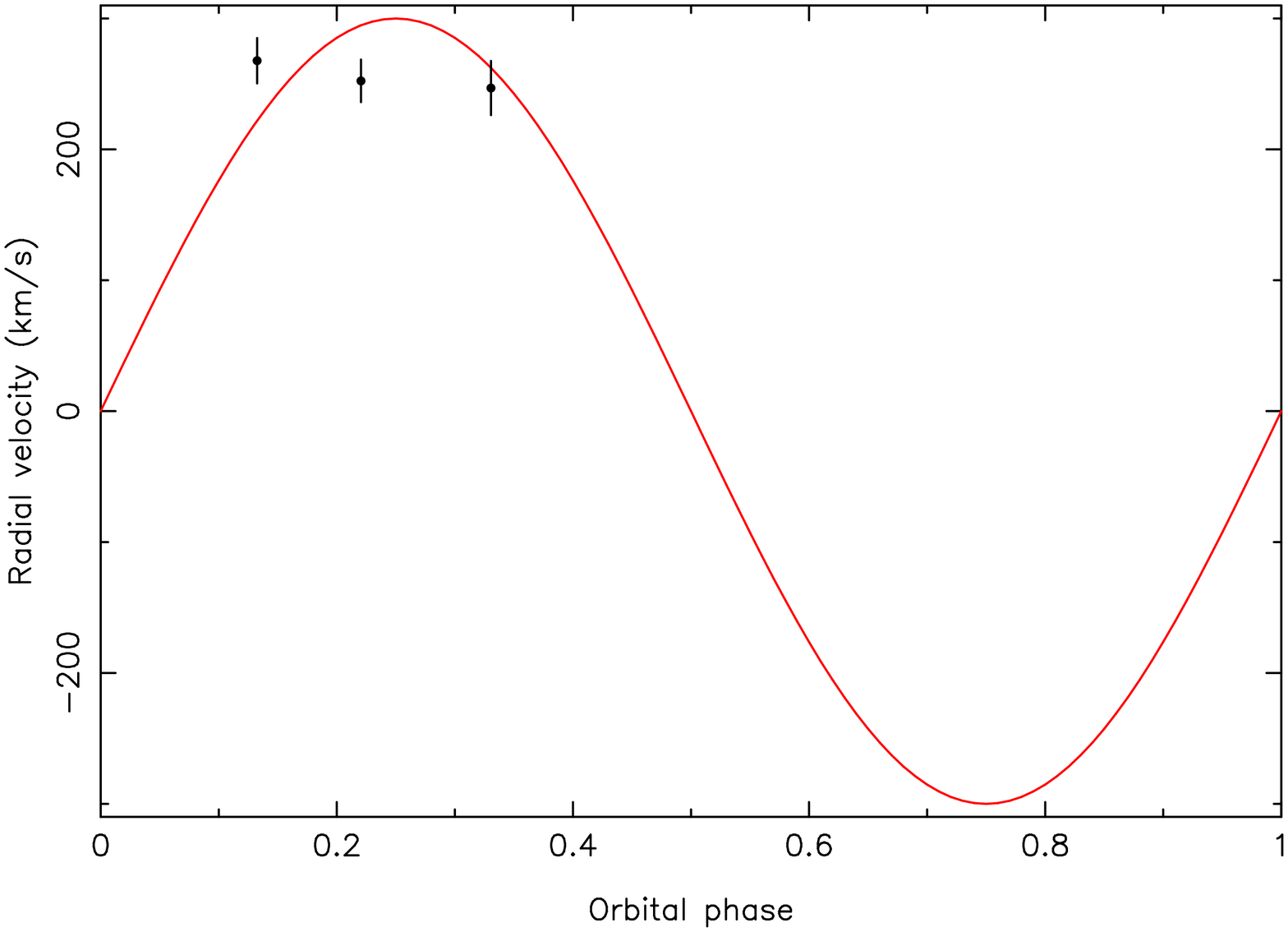}
 \hspace{2mm}
 \includegraphics[width=0.4\textwidth]{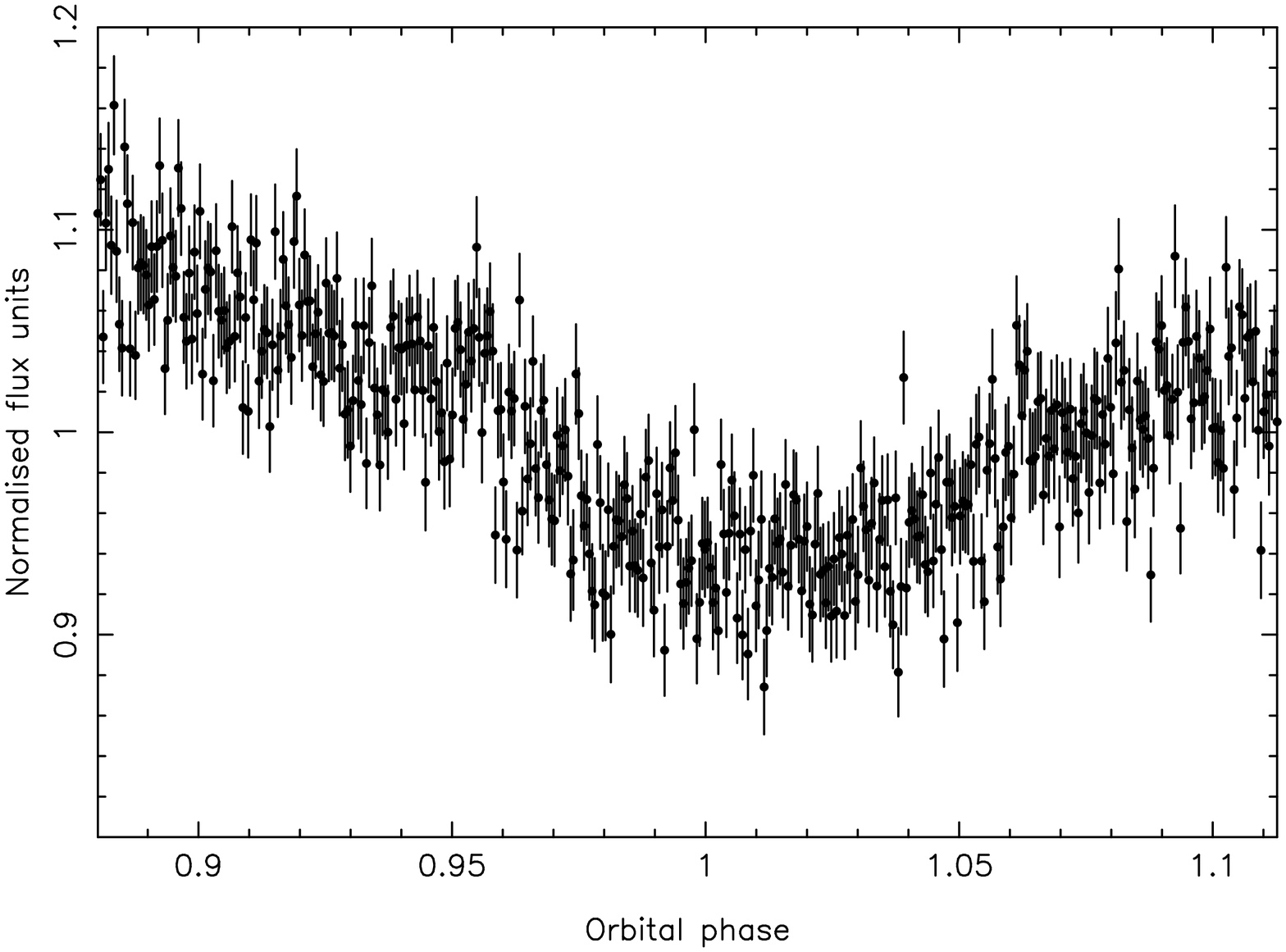}
 \caption{{\it Top left:} CSS light curve of \SDSS{0745+2631} showing large ellipsoidal modulation. The amplitude of this modulation is almost at the maximum possible value ($\sim0.3$ mags), implying that the main-sequence star is close to filling its Roche lobe. It also implies that the inclination of the system is high. {\it Top right:} SDSS spectrum of \SDSS{0745+2631} showing that the M star dominates the overall flux at optical wavelengths. {\it Bottom left:} Radial velocity measurements of the \Ion{Na}{i} 8200{\AA} doublet folded on the orbital period. Although these measurements cannot be used to measure the radial velocity amplitude of the M star they allowed us to determine which of the minima in the CSS light curve corresponded to phase zero (the eclipse of the white dwarf). {\it Bottom right:} follow up LT/RISE light curve of \SDSS{0745+2631} around the orbital phase of the putative white dwarf eclipse. The light curve appears to show a shallow eclipse-like feature superimposed on top of the ellipsoidal modulation. However, small flares from the M star could cause a similar feature and we are unable to say with certainty that the system is eclipsing. Photometry at shorter wavelengths, where the contribution from the white dwarf is larger, and hence the eclipse deeper, will prove if this system is eclipsing or not.}
 \label{fig:sdss0745_lcurves}
 \end{center}
\end{figure*}

\SDSS{0745+2631} was classified as a WDMS binary due to a slight blue excess, there is no spectroscopic evidence of a white dwarf in this system. The CSS light curve of this system shows very large ellipsoidal modulation, but only marginal evidence of an eclipse. The top-left panel of Figure~\ref{fig:sdss0745_lcurves} shows the CSS light curve of this system folded over its 5.2 hour period. The amplitude of this ellipsoidal modulation is related to the Roche lobe filling factor of the main-sequence star. For \SDSS{0745+2631} the amplitude is almost 0.3 magnitudes, which is the maximum possible value, implying that the main-sequence star almost fills its Roche lobe. The amplitude is also related to the orbital inclination, the large amplitude in this case implying that the inclination is high. Given that the flux of this system is dominated by the main-sequence star at visible wavelengths, with no white dwarf features seen in the SDSS spectrum (top-right panel of Figure~\ref{fig:sdss0745_lcurves}). Therefore, we would expect any eclipse to be shallow, and may be beyond the precision of the CSS data.

We determined which of the minima in the CSS light curve of \SDSS{0745+2631} corresponded to phase zero (the putative eclipse of the white dwarf) using the radial velocity measurements from the SDSS sub-spectra. In this case there were only three measurements of the \Ion{Na}{i} 8200{\AA} doublet all occuring at a similar orbital phase, with values $\sim250$\,\kms (bottom-left panel of Figure~\ref{fig:sdss0745_lcurves}). These observations were obtained near quadrature (either phase 0.25 or 0.75) and hence must have been taken at phase 0.25 (since at phase 0.75 we would expect the lines to be blueshifted rather than redshifted), which allowed us to determine which minima corresponded to phase zero.

Our follow-up RISE light curve is shown in the lower-right panel of Figure~\ref{fig:sdss0745_lcurves}. There is some evidence of a shallow dip around phase zero. However, the rapidly rotating M2 star is likely to be active and hence we would expect it to flare occasionally. A couple of unfortunately timed flares would give the same shape as a shallow eclipse meaning that, from this one observation alone, we cannot confirm the eclipsing nature of this binary. Observations at bluer wavelengths, where the contribution from the white dwarf is larger, may reveal the eclipse.

\subsubsection*{SDSS J082145.27+455923.4}

With a temperature of 80,000\,K, the 0.66\,{\MSUN} white dwarf in \SDSS{0821+4559} is the hottest white dwarf known in an eclipsing PCEB and a reflection component is easily visible in the CSS light curve. The binary has a fairly long period of 12.2 hours. It also contains a main-sequence star with one of the earliest spectral types, M2, in an eclipsing PCEB. The SDSS spectrum is dominated by the hot white dwarf. The \Ion{Na}{i} 8200{\AA} absorption doublet from the M star is also tentatively detected.

\subsubsection*{SDSS J092741.73+332959.1}

\SDSS{0927+3329} is the longest period eclipsing white dwarf binary currently known, with a period of 2.3 days (55.4 hours). The 0.59\,{\MSUN} white dwarf is relatively hot but no out-of-eclipse variations are seen in the CSS data. Both components are easily detected in the SDSS spectrum. 

\subsubsection*{SDSS J094634.49+203003.4}

\SDSS{0946+2030} contains a relatively cool 0.62\,{\MSUN} white dwarf with an M5 main-sequence companion in a 6.1 hour binary. Little variation is seen in the light curve outside of the eclipse. Both components are well detected in the SDSS spectrum.

\begin{figure*}
\begin{center}
 \includegraphics[width=0.325\textwidth]{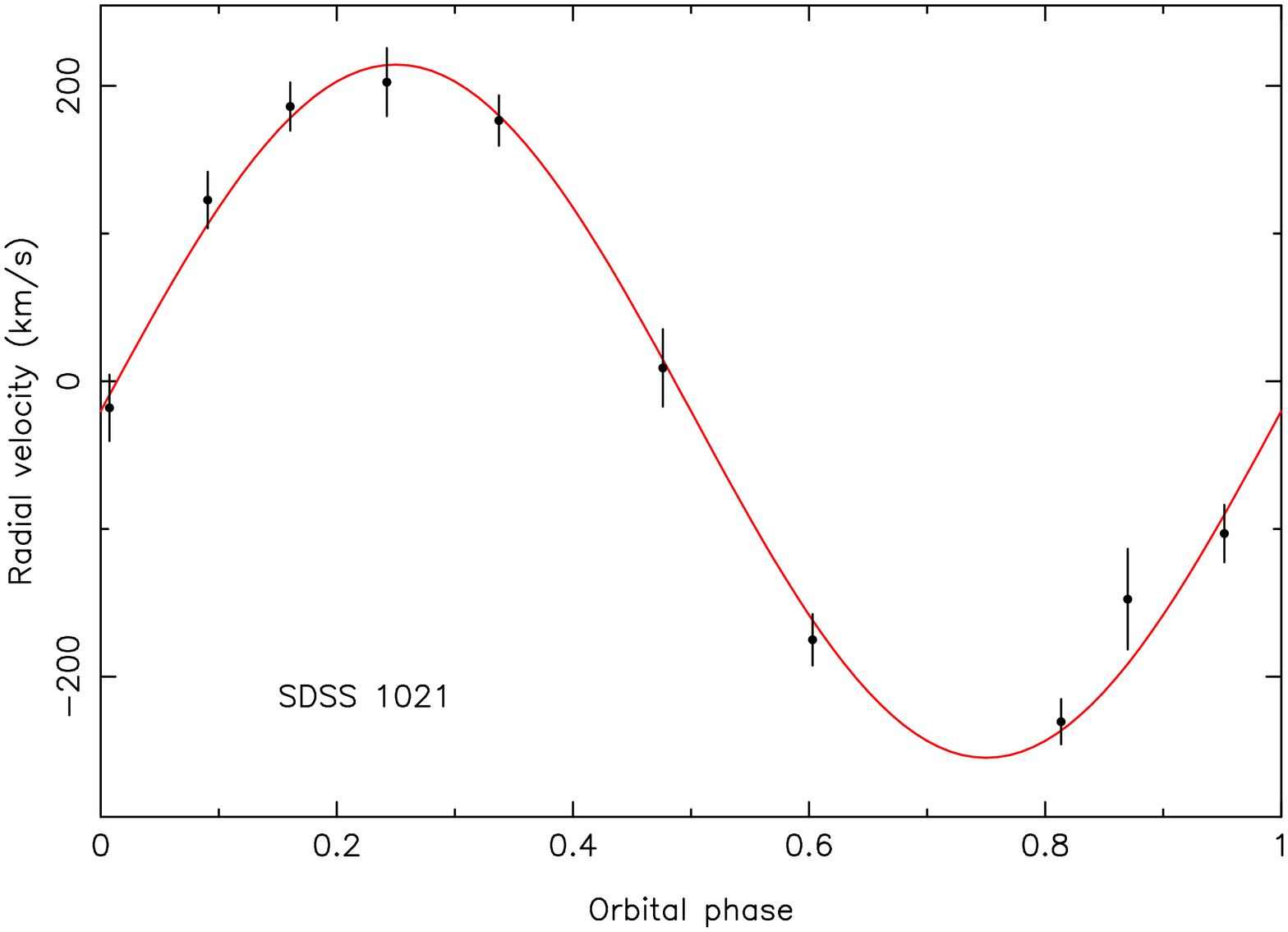}
 \includegraphics[width=0.325\textwidth]{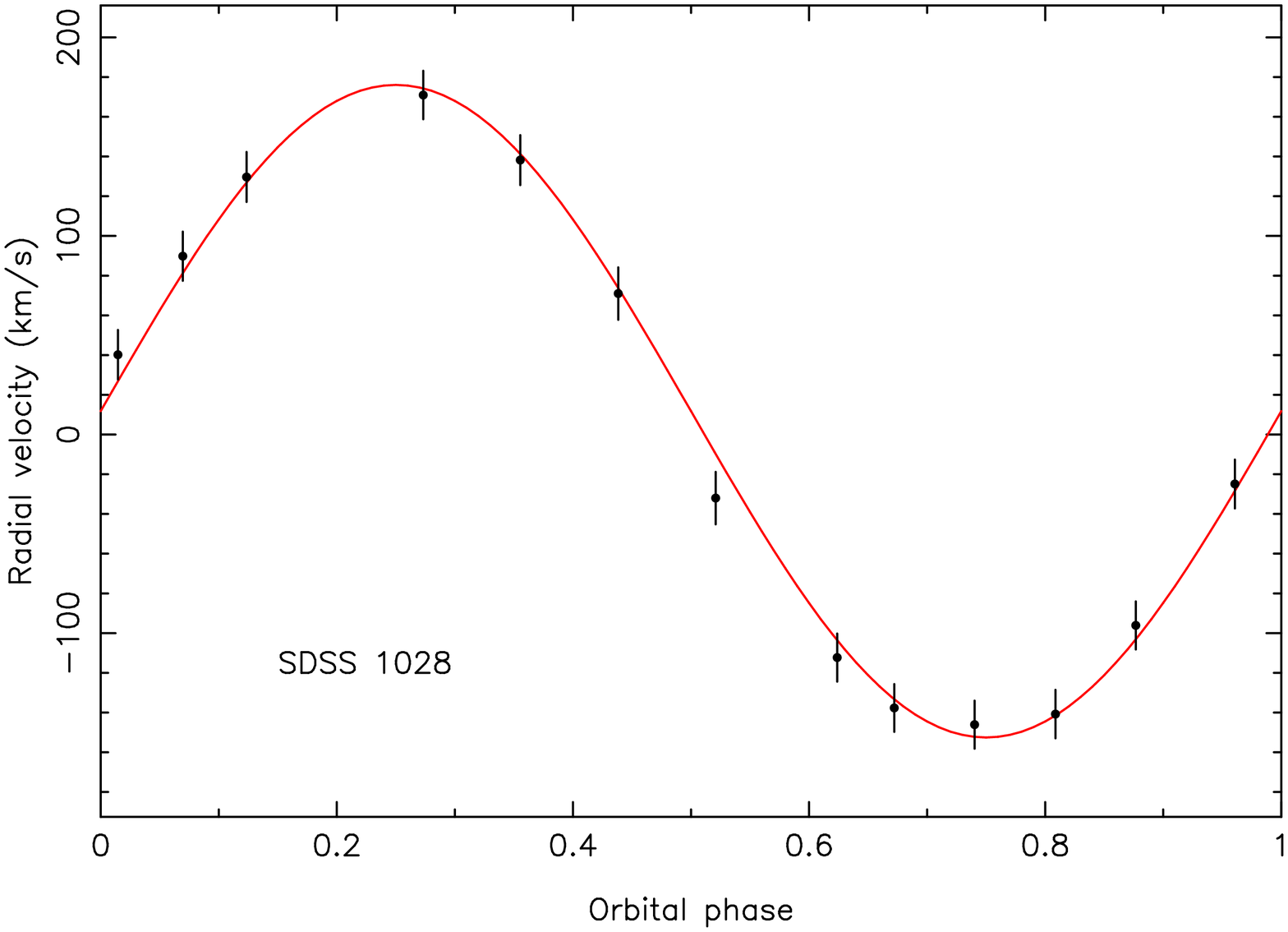}
 \includegraphics[width=0.325\textwidth]{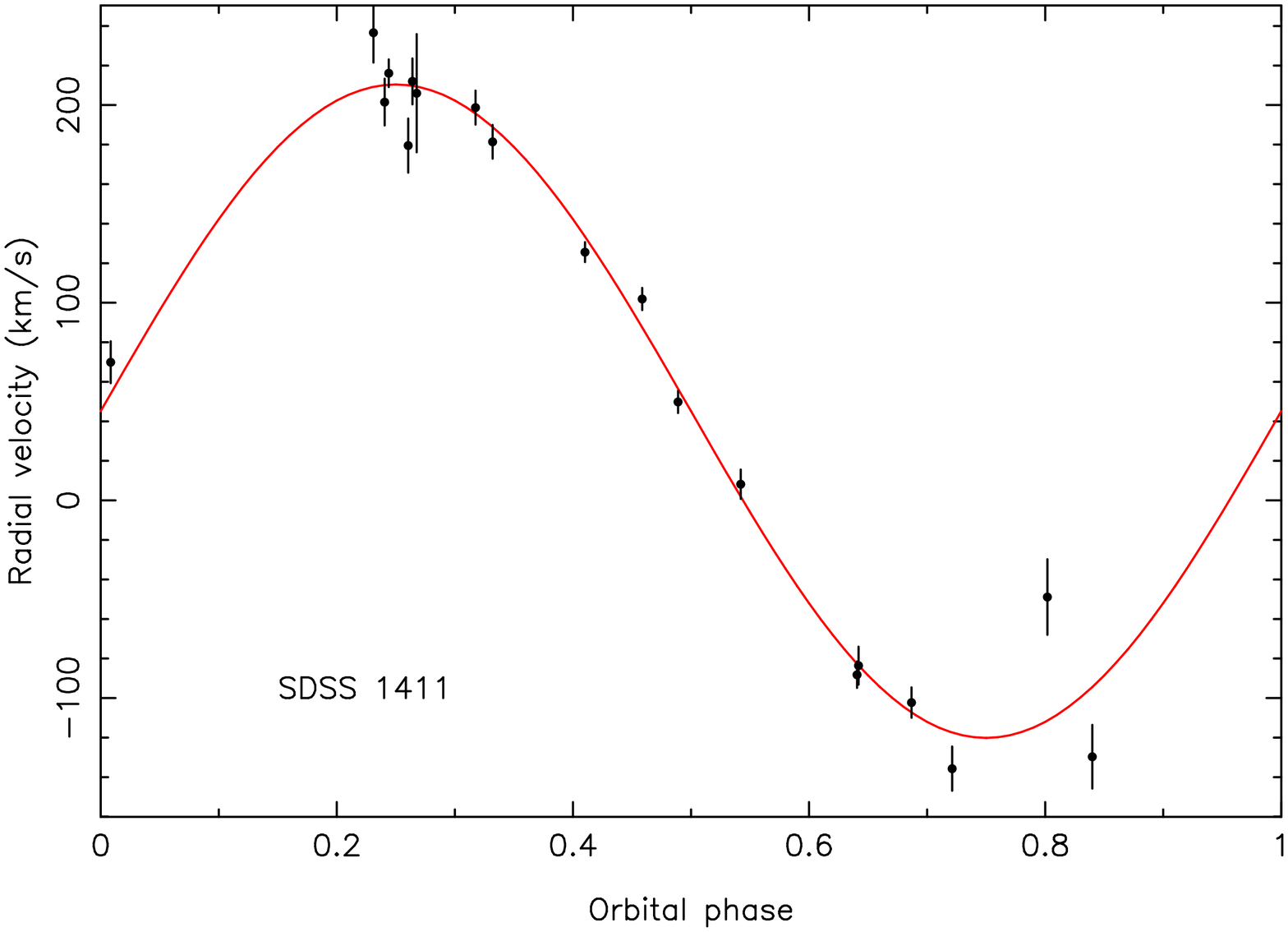}\\
 \vspace{1mm}
 \includegraphics[width=0.325\textwidth]{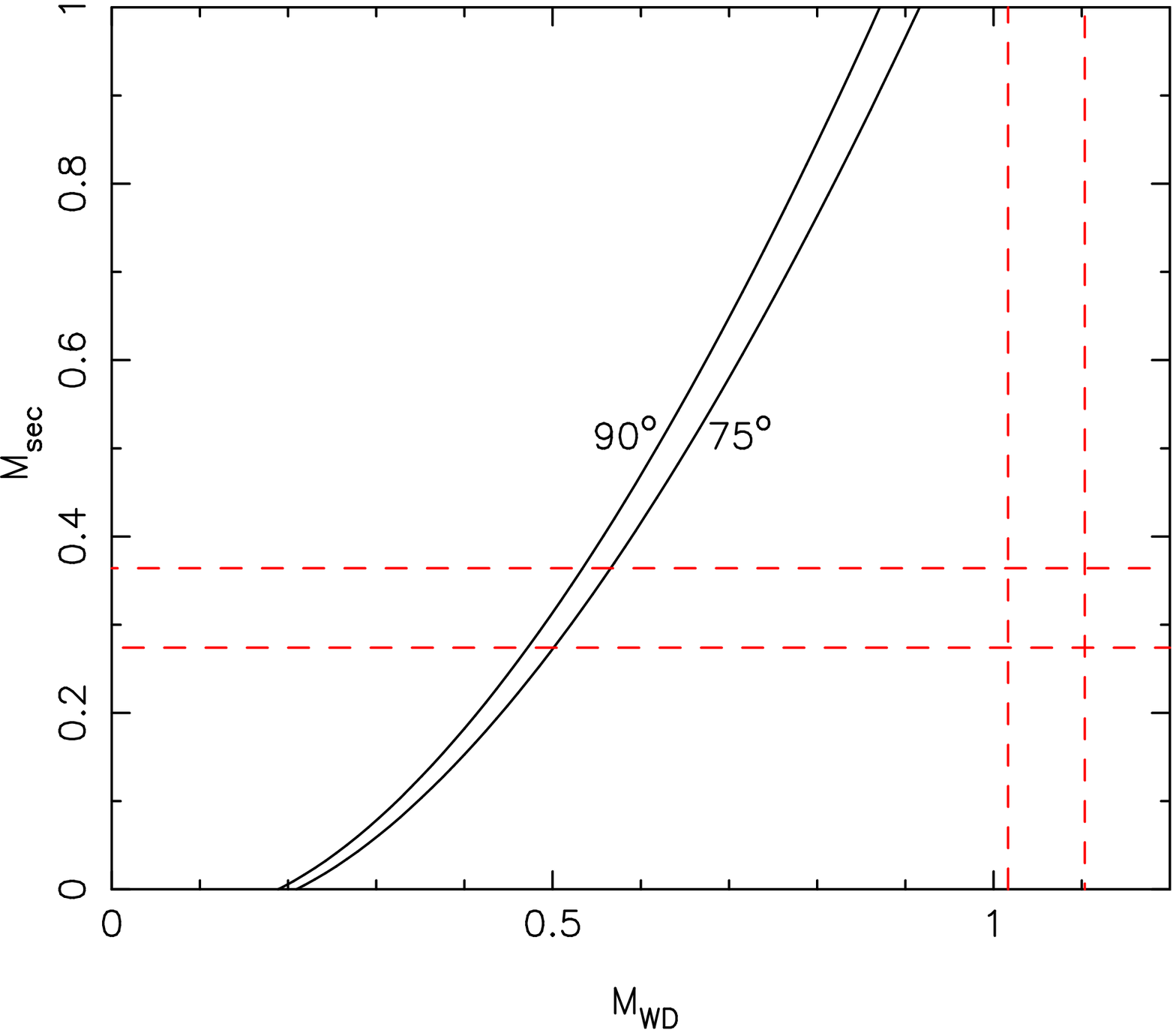}
 \includegraphics[width=0.325\textwidth]{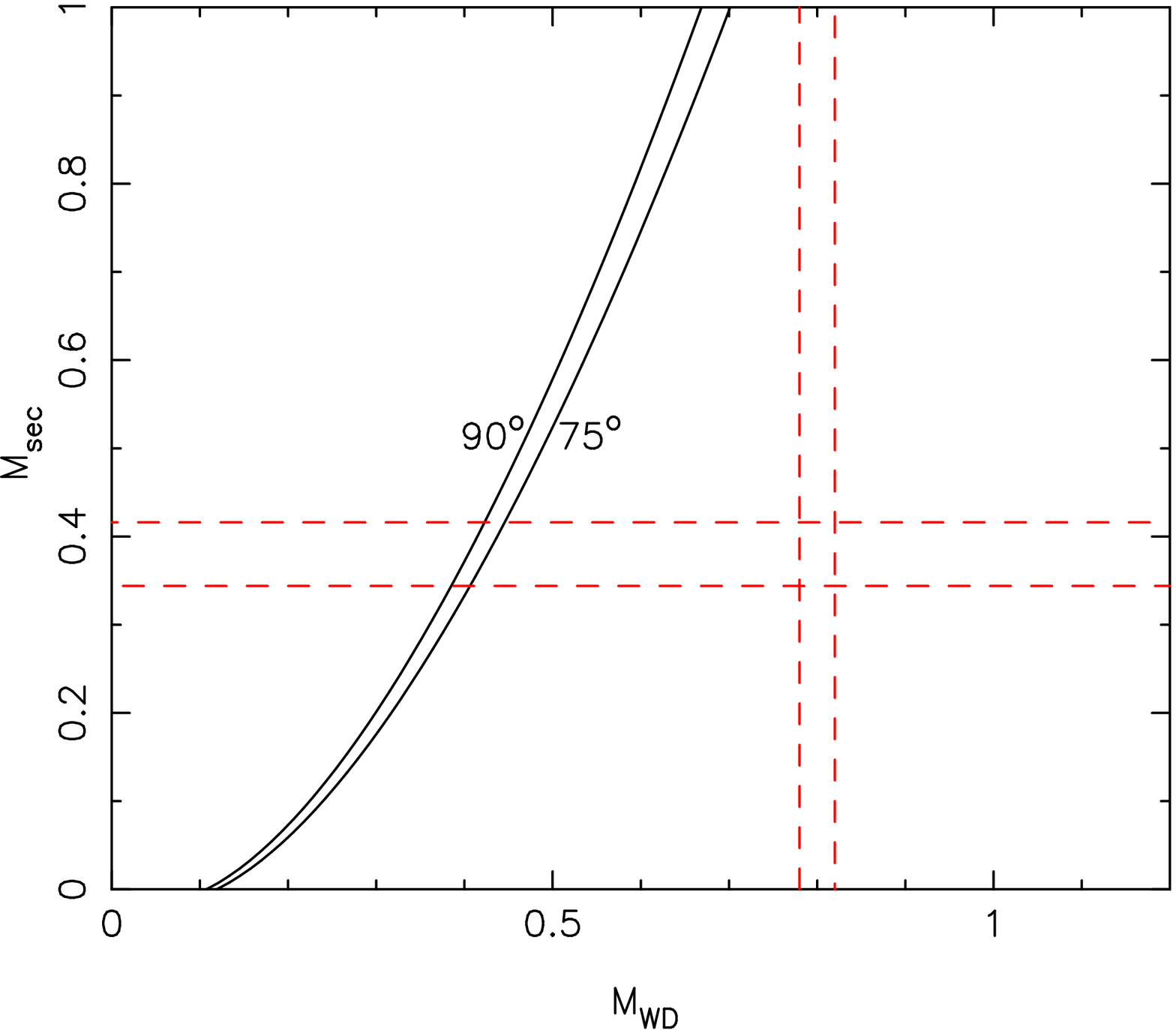}
 \includegraphics[width=0.325\textwidth]{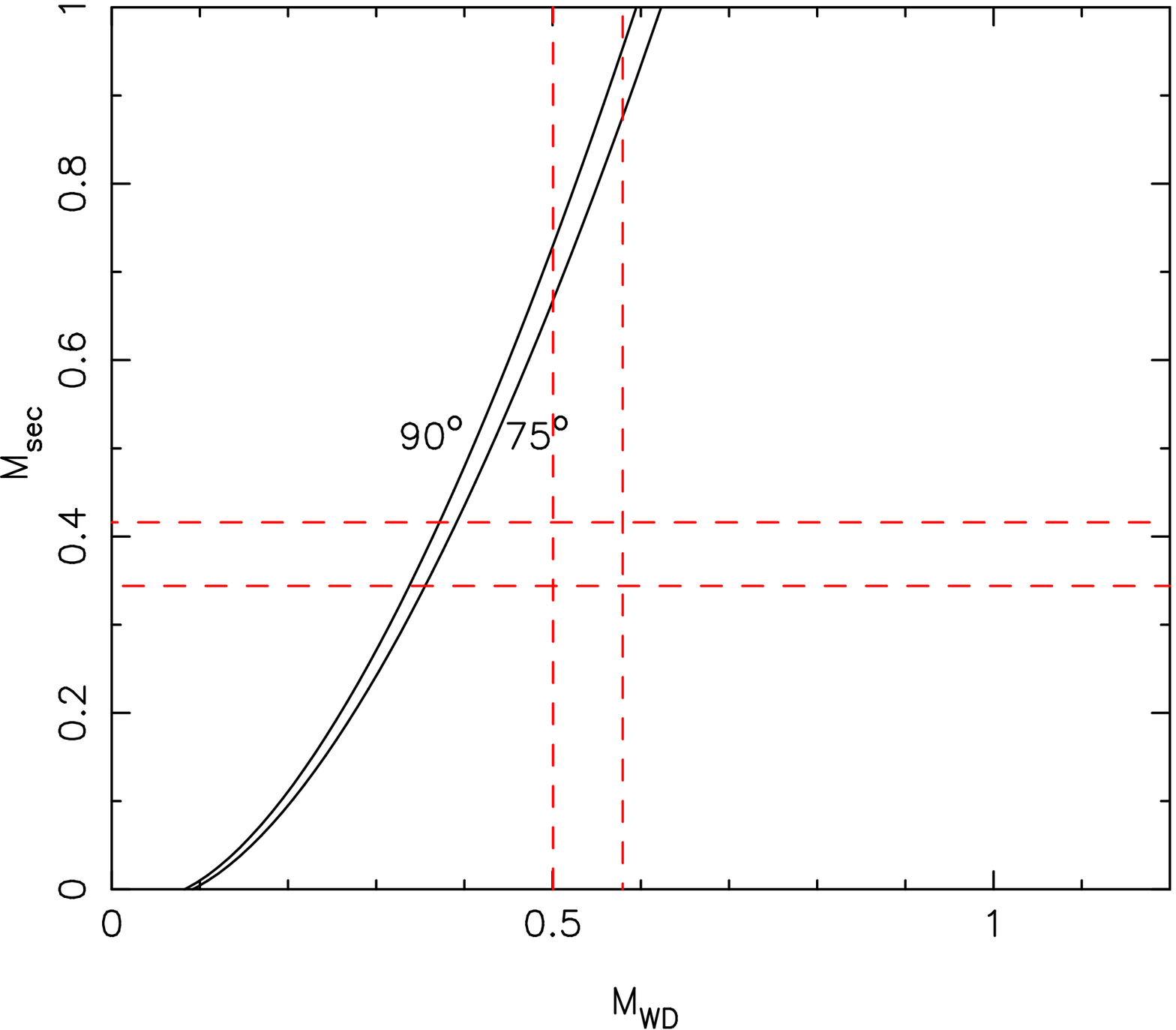}
 \caption{Radial velocity curves (top) and mass function plots (bottom) for \SDSS{1021+1744} (left), \SDSS{1028+0931} (centre) and \SDSS{1411+1028} (right). The dashed lines on the mass function plots indicate the limits on the mass of the main-sequence star (horizontal) and white dwarf (vertical) based on the spectral deconvolution. In all cases the implied mass of the white dwarf is lower than that determined from the spectral deconvolution.}
 \label{fig:mfunc}
 \end{center}
\end{figure*}

\subsubsection*{SDSS J095737.59+300136.5}

Its period of 1.9 days (46.2 hours) means that \SDSS{0957+3001} has the second longest period of all known eclipsing PCEBs, behind \SDSS{0927+3339}. Unsurprisingly, there is no evidence of out-of-eclipse variations in the CSS data. Both components are visible in the SDSS spectrum. The hot white dwarf has a fairly low mass of 0.42\,{\MSUN}, and is hence likely to have a helium core.

\subsubsection*{SDSS J102102.25+174439.9}

The CSS light curve of \SDSS{1021+1744} is dominated by the main-sequence star and there is a clear ellipsoidal modulation component. Our follow-up LT/RISE photometry revealed a large dip in the brightness of the system $\sim$15 minutes after the end of the eclipse (see Figure~\ref{fig:followup}). There is also some evidence for a flare from the main-sequence star occuring during the egress of the white dwarf. It is possible that this dip is caused by material ejected during this flare (or a previous flare) passing in front of the white dwarf, similar features have been seen in the eclipsing PCEB QS Vir \citep{odonoghue03,parsons11}. If this is the case then a large amount of material must have been ejected, since almost half of the white dwarf's flux is blocked. 

The SDSS spectrum is in fact a composition of 10 sub-spectra. The radial velocity measurements of the \Ion{Na}{i} 8200{\AA} absorption doublet reported by \citet{rebassa12a} combined with our ephemeris allowed us to measure the radial velocity amplitude of the main-sequence star as $K_\mathrm{sec}=235\pm9$\,\kms, with a systemic velocity of $\gamma=-20\pm6$\,\kms (see Figure~\ref{fig:mfunc}). With this information, and the orbital period, we can constrain the mass of the white dwarf, using the mass function,
\begin{equation} \label{eqn:mfunc}
f(M_\mathrm{WD})=\frac{(M_\mathrm{WD} \sin{i})^3}{(M_\mathrm{WD}+M_\mathrm{sec})^2}=\frac{P_\mathrm{orb}K_\mathrm{sec}^3}{2 \pi G}.
\end{equation}
For a given inclination, $i$, Equation~\ref{eqn:mfunc} defines the relationship between $M_\mathrm{WD}$ and $M_\mathrm{sec}$. The lower-left hand panel of Figure~\ref{fig:mfunc} shows this relationship for \SDSS{1021+1744} for an inclination of $90^\circ$ and $75^\circ$, the probable range over which the system is eclipsing. Also shown are the limits on the masses of the two stars from the deconvolution of the SDSS spectrum. It is clear that the mass of the white dwarf from the deconvolution ($1.06\pm0.09$\MSUN) is a substantial overestimate since the main-sequence star would have to have a mass in excess of 1\MSUN, certainly not an M dwarf. Assuming that the constraint on the mass of the secondary star is correct, the mass of the white dwarf is closer to $0.5$\MSUN. This discrepancy could be caused by the fact that this system is faint and the signal-to-noise of the SDSS spectrum is low. Furthermore, the main-sequence star dominates the spectrum hence the fit to the few white dwarf features visible is relatively poor.

\subsubsection*{SDSS J102857.78+093129.8}

\SDSS{1028+0931} is the brightest of our new eclipsing systems and has a period of 5.6 hours. The flux is dominated by the main-sequence star at visible wavelengths, although the white dwarf's features are still visible in the SDSS spectrum. The CSS light curve shows evidence of ellipsoidal modulation. There are 13 SDSS sub-spectra for this object. \citet{rebassa12a} measured the radial velocity of the main-sequence star from the \Ion{Na}{i} 8200{\AA} absorption doublet for each of these sub-spectra. Using these measurements and our ephemeris we were able to determine that the radial velocity amplitude of main-sequence star is $K_\mathrm{sec}=164\pm5$\,\kms, with a systemic velocity of $\gamma=12\pm4$\,\kms (upper-centre panel of Figure~\ref{fig:mfunc}). 

As with \SDSS{1021+1744}, we can use the $K_\mathrm{sec}$ measurement to constrain the mass of the white dwarf. The lower-centre panel of Figure~\ref{fig:mfunc} shows the relationship between $M_\mathrm{WD}$ and $M_\mathrm{sec}$ for high inclinations. Like \SDSS{1021+1744} the mass of the white dwarf from the deconvolution ($0.80\pm0.04$\MSUN) is a substantial overestimate. Assuming that the constraint on the mass of the secondary star is correct, the mass of the white dwarf roughly $0.42$\MSUN. However, like \SDSS{1021+1744}, the SDSS spectrum is dominated by the main-sequence star to such an extent that it may have affected the fit to the white dwarf features.

\subsubsection*{SDSS J105756.93+130703.5}

The mass of the white dwarf in \SDSS{1057+1307} determined from the spectral decomposition of the SDSS spectrum is 0.34\,{\MSUN} \citep{rebassa10}, making it the lowest mass white dwarf in all our new eclipsing systems and likely to have a helium core. The SDSS spectrum is dominated by the white dwarf but there are some features from the main-sequence star at long ($>7000$\AA) wavelengths. The orbital period is almost exactly 3 hours.

\subsubsection*{SDSS J122339.61-005631.1}

\SDSS{1223-0056} was observed with ULTRACAM mounted on the NTT as part of a project to detect pulsating white dwarfs in white dwarf plus main-sequence binaries. No pulsations were seen, but an eclipse was recorded. Therefore, we knew in advance that we might see eclipses in the CSS light curve of this system. \SDSS{1223-0056} is a partially eclipsing system and hence the eclipse only lasts around 5 minutes (see Figure~\ref{fig:followup}). Nevertheless, 3 CSS observations were taken in eclipse. The CSS light curve also shows some evidence of ellipsoidal modulation. This system has the shortest orbital period of all our newly discovered eclipsing systems, 2.1 hours. The white dwarf has a low mass of 0.45\,{\MSUN}, making it likely to be a helium core white dwarf. The main-sequence star has a spectral type of M6, making it the latest spectral type from all our new eclipsing systems.

\subsubsection*{SDSS J130733.49+215636.7}

The white dwarf in \SDSS{1307+2156} is a DC white dwarf and hence has a featureless spectrum. Therefore, we have no information on its mass. The temperature of the white dwarf is limited to $<$8000\,K based on the lack of Balmer absorption lines and a Galaxy Evolution Explorer (GALEX) near-UV magnitude of $21.05\pm0.22$. Our follow up LT/RISE photometry of the eclipse revealed a very sharp ingress and egress, lasting $\sim25$ seconds each (see Figure~\ref{fig:followup}). The short duration of these features implies that the white dwarf is quite small and is therefore likely to be quite massive. However, radial velocity information is needed in order to constrain the white dwarf's mass. The CSS light curve also shows evidence of ellipsoidal modulation over the 5.2 hour orbital period.

\SDSS{1307+2156} is only the second known eclipsing non-DA white dwarf after \SDSS{0303+0054} \citep{pyrzas09,debes12}. There are 5 other non-DA white dwarfs in (non-eclipsing) PCEBs \citep{nebot11}. Interestingly, all these white dwarfs are featureless DC white dwarfs, there is currently no known DB white dwarf in a PCEB\footnote{\citet{raymond03} claim to observe a 150\,{\kms} radial velocity variation in the DB+MS binary SDSS\,J144258.47+001031.5, based on 2 measurements. However, no variations are seen in the 18 radial velocity measurements listed in \citet{rebassa12a}, hence we conclude that the measurements of \citet{raymond03} are erroneous and this system is in fact a wide binary.}. This deficit is significant because 27 WDMS systems with DB white dwarfs have been spectroscopically followed up in order to determine whether they are close PCEBs \citep{rebassa12a,nebot11}. None of these showed any radial velocity variations, despite the fact that we would expect $\sim$$1/3$ to be PCEBs \citep{schreiber10,rebassa11}, hence this deficit appears to be genuine. 

The lack of DB white dwarfs in PCEBs is likely due to the white dwarf accreting some of the wind of its main-sequence companion. Wind accretion rates on to white dwarfs in PCEBs are of the order of $10^{-15}\msy$ \citep{debes06,tappert11,pyrzas12,parsons12b} meaning that the white dwarf will accrete $10^{-7}$\,{\MSUN} of hydrogen in 100\,Myr. The hydrogen in this wind will form a layer on the surface of the white dwarf, turning a DB white dwarf in to a DA white dwarf. However, DC white dwarfs are much cooler and have much deeper outer convection zones \citep{dufour07} which mixes the accreted hydrogen to such a low level that it is invisible. Hence we would still expect to see DC white dwarfs in PCEBs, as we do.

\subsubsection*{SDSS J140847.14+295044.9}

The 0.49\,{\MSUN} white dwarf in \SDSS{1408+2950} is relatively hot causing a small reflection effect, evident in the CSS light curve. It has an orbital period of 4.6 hours. The SDSS spectrum is dominated by the white dwarf, however, a large emission component is visible in the H$\alpha$ line. This emission is much stronger than in any of the other new eclipsing systems. Some of the main-sequence stars in the other systems are more highly irradiated than the one in \SDSS{1408+2950} meaning that this emission could also be due to activity on the main-sequence star.

\begin{table*}
 \centering
 \begin{minipage}{\textwidth}
 \renewcommand{\thempfootnote}{\fnsymbol{mpfootnote}}
  \centering
  \caption{Non-eclipsing PCEBs identified from the CSS photometry. Systems were identified using either a reflection effect (R) or ellipsoidal modulation (E). Stellar parameters taken from \citet{rebassa12a}} 
  \label{tab:nonecl}
  \begin{tabular}{@{}llllllll@{}}
    \hline
    SDSS Name & WD mass  & WD $T_\mathrm{eff}$ & Sp type of  & $r$ mag & Period & Type & Amplitude \\
              & (\MSUN)  & (K)               &  MS star    &         & (days) &      & (mags)    \\
    \hline
    SDSS\,J074807.22+205814.2 & $0.52\pm0.06$ & $86726\pm7788$ & M2.0 & 18.55 & 0.07205455(1) & R & 0.035 \\
    SDSS\,J080304.61+121810.3 & $1.04\pm0.27$ & $15071\pm2261$ & M2.0 & 17.14 & 0.5723126(29) & E & 0.023 \\
    SDSS\,J083618.61+432651.5 & $0.46\pm0.03$ & $24726\pm521$  & M3.0 & 17.98 & 0.19689803(95)& R & 0.049 \\
    SDSS\,J091211.01+442057.8 &               &                & M0.0 & 17.30 & 0.7311376(50) & E & 0.052 \\
    SDSS\,J091216.37+234442.5 & $0.69\pm0.03$ & $30071\pm245$  & M3.0 & 17.66 & 0.2635582(5)  & R & 0.065 \\
    SDSS\,J113316.27+270747.6 & $0.56\pm0.05$ & $72971\pm4638$ & M1.0 & 18.00 & 0.1781628(12) & R & 0.023 \\
    SDSS\,J114509.77+381329.2 &               &                & M4.0 & 15.95 & 0.19003799(27)& E & 0.027 \\
    SDSS\,J115857.33+152921.4 & $0.80\pm0.10$ & $36996\pm1203$ & M3.0 & 18.86 & 0.06666328(14)& R & 0.053 \\
    SDSS\,J122630.86+303852.5 & $0.40\pm0.01$ & $30071\pm66$   & M3.0 & 16.41 & 0.2586905(9)  & R & 0.061 \\
    SDSS\,J122930.65+263050.4 & $1.04\pm0.08$ & $21045\pm820$  & M3.0 & 17.30 & 0.6711480(66) & E & 0.045 \\
    SDSS\,J155904.62+035623.4\footnote[2]{Found by \citet{nebot11}} & $0.68\pm0.09$ & $48212\pm2446$ & & 18.58 & 0.0943473(1)  & R & 0.120 \\
    SDSS\,J162558.25+351035.7 &               &                & K7.0 & 17.40 & 0.3856815(25) & R & 0.062 \\
    SDSS\,J173002.48+333401.8 & $0.44\pm0.03$ & $47114\pm1176$ &      & 18.39 & 0.1569473(3)  & R & 0.140 \\
\hline
\end{tabular}
\vspace{-6mm}
\end{minipage}
\end{table*}

\subsubsection*{SDSS J141134.70+102839.7}

\citet{tappert11} first presented evidence that \SDSS{1411+1028} was a PCEB. \citet{nebot11} determined its period as 4.0 hours and measured the radial velocity amplitude of main-sequence star as $K_\mathrm{sec}=168\pm4$\,\kms. Due to its eclipsing nature, our CSS photometry gives tighter constraints on its period. We also detect a reflection component in the out-of-eclipse light curve. The spectrum is dominated by the relatively hot white dwarf.

Since we have a measurement of the radial velocity amplitude of the main-sequence star we can use it to constrain the mass of the white dwarf. The lower-right hand panel of Figure~\ref{fig:mfunc} shows the relationship between $M_\mathrm{WD}$ and $M_\mathrm{sec}$ for high inclinations. As with the two other systems where we have radial velocity information, the mass of the white dwarf determined from the deconvolution of the SDSS spectrum ($0.54\pm0.08$\MSUN) is an overestimate, although the discrepancy is smaller in this case. Assuming that the constraint on the mass of the secondary star is correct, the mass of the white dwarf roughly $0.36$\MSUN, making it a firm helium core candidate. In this case the low signal-to-noise of the SDSS spectrum may have contributed to this overestimation.

\subsubsection*{SDSS J223530.61+142855.0}

Like \SDSS{1223-0056} we had prior knowledge that \SDSS{2235+1428} was an eclipsing system. The radial velocity of the main-sequence star was observed to change by almost 500\,{\kms} between two nights, implying that the system was not only a PCEB, but that it was also a high inclination system. Subsequent photometric follow up revealed that the system was eclipsing and that its period was 3.4 hours. The deep white dwarf eclipse is clearly visible in the CSS light curve. However, there is little out-of-eclipse variation. The SDSS spectrum is dominated by the white dwarf. The \Ion{Na}{i} 8200{\AA} absorption doublet from the main-sequence star is just visible. With a mass of 0.45\,{\MSUN} the white dwarf in \SDSS{2235+1428} is another helium core candidate. 

\section{Non-eclipsing PCEBs}

As noted in Section~\ref{sec:eclipsers}, we performed a period search on all of our light curves to search for shallower eclipses. However, this approach also revealed the periods of several non-eclipsing systems via reflection or ellipsoidal modulation effects. 

Since these effects can be quite small we used 4 different period analyses: Scargle \citep{scargle82}, a straight power spectrum, analysis of variance (AoV, \citealt{schwarzenberg89}) and ANoVA \citep{schwarzenberg96}. We consider the period robust if all 4 methods give, within one harmonic, the same period. 

Using photometry to identify and determine the periods of PCEBs is not particularly efficient. A much better approach is to try to detect radial velocity variations, since these are easier to detect and much less biased towards hot white dwarfs, short periods and large Roche-lobe filling factors. This approach has already led to the discovery of dozens of PCEBs \citep{rebassa07,rebassa10,rebassa11,schreiber08,schreiber10,nebot11}. Nevertheless, using the CSS photometry we were able to determine the periods of 13 PCEBs, including the previously know PCEB \SDSS{1559+0356}, for which we measure the same period as \citet{nebot11}. These non-eclipsing systems are detailed in Table~\ref{tab:nonecl}. 

\section{Discussion}

\subsection{The percentage of eclipsing systems}

Among our 835 objects, roughly one third ($\sim$$280$) will be close PCEBs \citep{schreiber10}. Of these 29 are eclipsing ($\sim$$10\%$). To test if this relatively high percentage is consistent with our current knowledge of PCEBs we simulated the light curves of a population of PCEBs with random inclinations and measured the percentage that were eclipsing. We used the period distribution of \citet{nebot11} and the white dwarf mass distribution of \citet{zorotovic11a}. For the main-sequence star we adopted the mass distribution of \citet{zorotovic11b} and included their correlation between the orbital period and main-sequence star mass. The combination of these distributions provided the two stellar masses and the orbital period. We then used the mass-radius relationship for white dwarfs of Eggleton, quoted in \citet{verbunt88} and the mass-radius relationship for a 3\,Gyr main-sequence star from \citet{baraffe98}. We accounted for Roche distortion and reject any systems in which  the secondary star exceeded its Roche lobe, then tested if the system was eclipsing. 

We found that $12\%$ of the simulated PCEBs were eclipsing, consistent with the number found in the CSS photometry. We also simulated populations with different mass distributions and found that the number of eclipsing systems is very insensitive to the mass of the white dwarf, due to its small size. However, the number of eclipsing systems is quite dependent upon the mass of the main-sequence star, with a larger number of systems eclipsing with more massive (and hence larger) main-sequence stars. For example, scaling the mass distribution of \citet{zorotovic11b} to peak at 0.4$\MSUN$ increases the percentage of eclipsing systems to $15\%$.

\subsection{Completeness} \label{sec:completeness}

\begin{figure}
\begin{center}
 \includegraphics[width=\columnwidth]{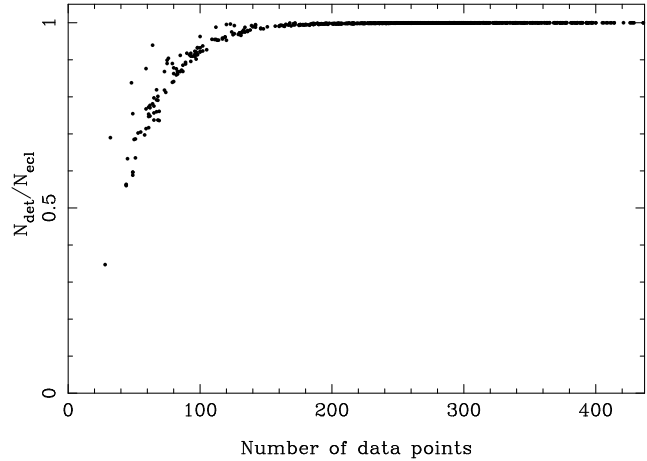}
 \caption{Probability of detecting eclipsing PCEBs based on the number of CSS data points. Each point represents the temporal sampling of a CSS light curve.}
 \label{fig:ecl_detect}
 \end{center}
\end{figure}

We used our simulated PCEB light curves to test our eclipse detection completeness. For each CSS light curve in our sample, we took the temporal sampling and created 100,000 synthetic PCEB light curves and tested how many of the eclipsing systems were detected. We classified a system as a confirmed eclipsing system if 2 or more separate eclipses were detected. The results of this are shown in Figure~\ref{fig:ecl_detect}. We found that for light curves with 100 data points $90\%$ of eclipsing systems were detected, and with 200 data points $99\%$ of eclipsing systems were detected. Only for light curves that have less than 100 points is there a reasonable chance of missing some eclipsing systems (these missed systems are usually systems with periods in excess of 1 day). Since the majority of our light curves comprise of more than 200 observations ($76\%$) our overall detection percentage is $\sim$$97\%$. Therefore it is very unlikely that the well-sampled light curves that show no eclipses are in fact eclipsers (and that we have just missed the eclipses). 

This simulation did not take into account our ability to visually detect the eclipse in the CSS data. There could potentially be some very shallow eclipsing systems which we would not detect due to the signal-to-noise of the CSS data (e.g. systems with very cool white dwarfs). Therefore, our completeness calculations should be viewed as upper limits.

\subsection{Long period systems}

Our sample of new eclipsing systems contain two systems with periods in excess of 1.9 days. These systems are particularly well suited for studying any long term orbital period variations, which in some cases have been attributed to the presence of planets in orbit around the binary \citep{beuermann10,parsons10b,kaminski07,potter11,beuermann11}. However, the effects of quadrupole moment fluctuations driven by stellar activity cycles, known as the Applegate mechanism \citep{applegate92}, can complicate the analysis of eclipse time variations by adding noise on potentially the same scale as that caused by any third body. 

The amount of energy required to drive period changes via Applegate's mechanism scales as $(a/R)^2$, where $a$ is the orbital separation and $R$ is the radius of the main-sequence star \citep{applegate92}. Therefore, any period variations caused by Applegate's mechanism in these longer period systems will be negligible. This makes these systems ideal to search for planets since Applegate's mechanism can be ruled out as the cause of any observed period variations, leaving few alternative explanations other than the reflex motion of the binary caused by an unseen body in orbit around them.

\subsection{Future Evolution}

Detached PCEBs are the direct progenitors of cataclysmic variables (CVs). As such they may provide crucial information for our understanding of CV evolution because, in contrast to CVs, it is possible to determine the evolutionary state of each PCEB using the temperature of the white dwarf which provide us with a robust age estimate. Furthermore, for a given angular momentum loss prescription, one can easily reconstruct the history of the systems and predict their future. 

Following \citet{schreiber03} and \citet{zorotovic11a} we search for PCEBs representative for the progenitors of the current CV population among the 11 new eclipsing PCEBs in our sample with reasonable estimates of the stellar masses and WD temperature (the only system we excluded is the DC white dwarf system \SDSS{1307+2156}). We require the CV formation time to be shorter than the Hubble time and the mass ratio $q=M_\mathrm{sec}/M_\mathrm{WD}$ to be $<0.67$ as otherwise the (second) mass transfer will be dynamically unstable. With this restriction only three PCEBs within our sample can be considered representative for the progenitors of the current CV population: \SDSS{0821+4559}, \SDSS{0946+2030} and \SDSS{1223-0056}. \SDSS{0821+4559} is one of the few known progenitors of CVs that will clearly start mass transfer above the orbital period gap ($P_\mathrm{orb}=3.94$\,hrs), while \SDSS{0946+2030} will become a CV in the gap ($P_\mathrm{orb}=2.32$\,hrs). The interpretation of \SDSS{1223-0056} as being a CV progenitor is ambiguous, as this particular system could also be a detached CV evolving through the gap \citep{davis08}. This second possibility appears to be very reasonable as the system is close to Roche-lobe filling and will start mass transfer in only $\sim\,2.7$\,Myrs, close to the lower boundary of the period gap at $P_\mathrm{orb}=2.24$\,hrs. 

The remaining systems in our sample have either very long CV formation time scales (true for the remaining systems containing carbon-oxygen core white dwarfs), or the second mass transfer will be dynamically unstable (true for the remaining systems with helium core white dwarfs), or both which happens to be the case for the long orbital period PCEB containing a helium core primary (\SDSS{0957+3001}).  

Finally, we note that the above calculations should be taken as first order estimates mostly because of two systematic uncertainties. Firstly, the stellar masses were derived from fitting the SDSS spectrum which we have shown to be potentially quite rough (see Figure~\ref{fig:mfunc}). Secondly, we used an empirical spectral type-radius relation to estimate the radii of the secondary stars, the intrinsic scatter around any such relation is know to be significant (for details see \citealt{rebassa07}).

\section{Conclusions}

We have analysed the Catalina Sky Survey light curves of all white dwarf plus main-sequence binaries in the catalogue of \citet{rebassa12a} with $g<19$, in a search for new eclipsing systems. We identify a total of 29 eclipsing systems, 12 of which were previously unknown, and one candidate eclipsing system which needs better quality photometry for confirmation. This increases the number of known eclipsing post common envelope binaries to 49. We present high-speed follow-up light curves of all our newly identified systems, confirming both their eclipsing nature and their ephemerides.

We find two new eclipsing systems with periods in excess of 1.9 days. These systems are ideal targets for detecting planets in orbit around the binary via orbital period variations. This is because a common source of noise in the eclipse time variations, known as Applegate's effect, will have a reduced impact on the timing variations in these long period systems.

Our newly discovered systems cover a large variety of parameters. We find one system with a very hot white dwarf ($T_\mathrm{eff}=80938$\,K), a system with a featureless DC type white dwarf and a system with a very low mass white dwarf ($M_\mathrm{WD}\sim0.3$\MSUN). The main-sequence stars span spectral types from M2-M6.

For three systems we were able to place constraints on the mass of the white dwarf using measurements of the radial velocity amplitude of the main-sequence star. In all cases the mass is lower than implied from the deconvolution of the SDSS spectrum. Therefore, the system parameters of all our newly identified systems are subject to some uncertainty until more detailed studies \citep[e.g.][]{parsons10a,parsons12b,pyrzas12} are carried out on them.

\section*{Acknowledgments}

SGP acknowledges support from the Joint Committee ESO-Government of Chile. ULTRACAM, BTG, TRM, VSD and SPL are supported by the Science and Technology Facilities Council (STFC). ARM acknowledges financial support from  FONDECYT in the form of grant number  3110049. MRS thanks FONDECYT (project 1100782) and the Millennium Science Initiative, Chilean Ministry of Economy, Nucleus P10-022-F. The results presented in this paper are based on observations collected at the European Southern Observatory under programme IDs 086.D-0161 and 087.D-0557. The CSS survey is funded by the National Aeronautics and Space Administration under Grant No. NNG05GF22G issued through the Science Mission Directorate Near-Earth Objects Observations Program. The CRTS survey is supported by the U.S.~National Science Foundation under grants AST-0909182. The Liverpool Telescope is operated on the island of La Palma by Liverpool John Moores University in the Spanish Observatorio del Roque de los Muchachos of the Instituto de Astrofisica de Canarias with financial support from the UK Science and Technology Facilities Council. 

\bibliographystyle{mn_new}
\bibliography{crts_pcebs}

\label{lastpage}

\end{document}